%% file: approx_zero_fs.tex
\documentclass[11pt,a4paper]{article}
\pdfoutput=1

\usepackage{a4wide}
\usepackage{amsmath}
\usepackage{graphicx}

\DeclareMathOperator{\re}{Re}

\title{Investigation of the factorization scheme dependence of finite order perturbative
QCD calculations: searching for approximately ZERO factorization scheme}
\author{Karel Kolar,\\ Institute of Physics, Academy of Sciences of the Czech Republic,\\
Na Slovance 2, 182 21 Prague 8, Czech Republic,\\ kolark@fzu.cz}
\date{}

\begin{document}

\maketitle

\begin{abstract}
The possibility of an improvement of current NLO Monte Carlo
event generators by means of choosing a suitable factorization
scheme is studied. The optimal factorization scheme for combining
initial state parton showers and NLO hard scattering cross-sections
is the ZERO factorization scheme, in which all NLO splitting
functions vanish. However, it has turned out that the ZERO
factorization scheme has a limited range of practical applicability.
Hence, this paper is focused on searching for a factorization
scheme which is applicable at the NLO in the full range of $x$
relevant for QCD phenomenology and simultaneously close to
the ZERO factorization scheme (i.e.\ the corresponding NLO
splitting functions are close to zero).
\end{abstract}

\input{approx_zero_fs1.tex}
\input{approx_zero_fs2.tex}
\input{approx_zero_fs3.tex}

\section*{Acknowledgments}
The author would like to thank J.\ Ch\'yla for careful reading of the manuscript
and valuable suggestions. This work was supported by the projects LC527 of Ministry
of Education and AVOZ10100502 of the Academy of Sciences of the Czech Republic.

\end{document}

%% file: approx_zero_fs1.tex
\section{Introduction}

At present time many QCD cross-sections at parton level are known
at the NLO accuracy and necessary algorithms, e.g.\ \cite{potter,
schorner,webber,herwig,pythia}, for their incorporation in Monte
Carlo event generators have been developed. However, these
algorithms attach initial state parton showers that are
generated on the basis of the LO splitting functions to NLO
hard scattering cross-sections obtained in the standard $\overline{\rm MS}$
factorization scheme because no satisfactory algorithm for
generating parton showers at the NLO accuracy has been found so far.
The reasons why generating initial state parton showers at
the NLO accuracy is difficult in the $\overline{\rm MS}$ factorization
scheme are following. The NLO splitting functions no longer correspond
to basic QCD vertices, are expressed by much more complicated
formulae than the LO ones and are negative for some $x$, which
prevents us from using straightforward probabilistic interpretation,
which is crucial for Monte Carlo simulations.

Initial state parton showers induce the scale dependence of parton
distribution functions, which is described by the evolution
equations. Monte Carlo simulations of initial state parton showers
have some important advantages in comparison with the analytical
calculations represented by the evolution equations --- they take
into account transverse momenta, include the color description and
the partons radiated in initial state parton showers participate
in hadronization.

Since current NLO Monte Carlo event generators attach LO initial
state parton showers to NLO hard scattering cross-sections
corresponding to the $\overline{\rm MS}$ factorization scheme, they
cannot predict some quantities, e.g.\ spectra of transverse momenta,
at the NLO accuracy.\footnote{However, the NLO accuracy of some
predictions, like those for total cross-sections, is not disturbed
by using LO parton showers if the NLO parton distribution
functions are used in the simulation of the hard scattering
cross-section.} The deficiency lying in the combination
of LO initial state parton showers and NLO hard scattering
cross-sections calculated in the $\overline{\rm MS}$ factorization scheme
could be removed by using the ZERO factorization scheme, in which
all NLO splitting functions vanish, and therefore all NLO corrections
are included in hard scattering cross-sections. Attaching formally
LO initial state parton showers to NLO hard scattering cross-sections
is thus consistent if we use the ZERO factorization scheme. The change
of the factorization scheme employed in NLO Monte Carlo event
generators requires only to transform hard scattering cross-sections
from the standard $\overline{\rm MS}$ factorization scheme to a new one
and to determine parton distribution functions in the new factorization
scheme. The existing algorithms for parton showering and for
attaching parton showers to NLO cross-sections need not be changed.

The ZERO factorization scheme was studied in detail in \cite{kolar}.
Unfortunately, it has turned out that from the practical point of
view, the ZERO factorization scheme cannot be applied at the NLO
in the full range of $x$ needed for QCD phenomenology. Hence, it is
worth trying to find some factorization scheme which is applicable
without any restrictions and simultaneously sufficiently close to
the ZERO factorization scheme (i.e.\ the corresponding NLO splitting
functions are sufficiently close to zero). A factorization scheme
satisfying these conditions will be called an approximately ZERO
factorization scheme in the following. An approximately ZERO factorization
scheme does not allow the construction of fully consistent NLO Monte Carlo
event generators based on initial state parton showers that are taken formally
at the LO. But its exploitation in NLO Monte Carlo event generators still makes
sense because it can significantly reduce the deficiency caused by combining
LO initial state parton showers with NLO hard scattering cross-sections
calculated in the standard $\overline{\rm MS}$ factorization scheme.

This paper is focused on searching for an approximately ZERO factorization
scheme. This search is described in detail in the next section. The summary
and conclusion are presented in section \ref{prtsumandconcl}. The notation
used in this text is the same as that in \cite{kolar}.

%% file: approx_zero_fs2.tex
\section{Searching for approximately ZERO factorization scheme}

The NLO splitting functions should be as small as possible in an approximately
ZERO factorization scheme. If NLO splitting functions are small, then they
should have little influence on the evolution of parton distribution functions.
Hence, the evolution of parton distribution functions in any factorization
scheme that can be considered as an approximately ZERO factorization scheme
should be close to their evolution calculated with absolutely zero NLO
splitting functions. This fact can be exploited in searching for an
approximately ZERO factorization scheme. Since an approximately ZERO
factorization scheme is aimed at using for the description of proton
collisions, it is desirable to minimize the influence of NLO splitting
functions on the evolution of parton distribution functions in the case
of the proton.

To quantify the influence of NLO splitting functions on the evolution of parton
distribution functions in a factorization scheme FS, let us introduce a set of
auxiliary parton distribution functions $\mathbf{D}_0(x, M, {\rm FS}, M_{\rm S})$
that are defined as follows:
\begin{itemize}
  \item their evolution in the factorization scale $M$ is formally LO (i.e.\ the
   evolution is calculated with vanishing NLO splitting functions),
   independently of the factorization scheme FS,
  \item the initial condition of the evolution is given as\\
   $\mathbf{D}_0(x, M = M_{\rm S}, {\rm FS}, M_{\rm S}) = \mathbf{D}(x, M_{\rm S}, {\rm FS})$
\end{itemize}
where $\mathbf{D}(x, M, {\rm FS})$ represents the parton distribution functions
of the proton. The auxiliary ``zero'' parton distribution functions
$\mathbf{D}_0(x, M, {\rm FS}, M_{\rm S})$ are fully calculable from the parton
distribution functions $\mathbf{D}(x, M, {\rm FS})$. In any factorization scheme
FS that is close to the ZERO factorization scheme, the ``zero'' parton distribution
functions $\mathbf{D}_0(x, M, {\rm FS}, M_{\rm S})$ should be close to the parton
distribution functions $\mathbf{D}(x, M, {\rm FS})$ for all values of $x$ and $M$,
independently of the ``starting'' factorization scale $M_{\rm S}$, which specifies
the initial condition of the evolution of the ``zero'' parton distribution functions.

An approximately ZERO factorization scheme was searched by minimizing
the difference between $\mathbf{D}_0(x, M, {\rm FS}, M_{\rm S})$ and
$\mathbf{D}(x, M, {\rm FS})$. A small difference between
$\mathbf{D}_0(x, M, {\rm FS}, M_{\rm S})$ and $\mathbf{D}(x, M, {\rm FS})$
is a necessary, but not sufficient, condition for the factorization scheme FS
to be close to the ZERO factorization scheme. The reason why this condition was
used for searching for an approximately ZERO factorization scheme is that
it takes into account the fact that an approximately ZERO factorization scheme
is aimed at using for the description of proton collisions.

In \cite{kolar}, it has been shown that factorization schemes specified via
the corresponding NLO splitting functions, which can be chosen at will, may
have unexpected restrictions of their practical applicability --- NLO splitting
functions appearing at first sight as reasonable may specify a factorization
scheme that cannot be applied in the full range of $x$ relevant for QCD phenomenology.
Unexpected restrictions of practical applicability are ruled out if the Mellin
moments of the appropriate NLO splitting functions satisfy (see
\mbox{subsection 4.4} in \cite{kolar})
\begin{align}
  & P^{(0)}_{GQ}(n) \left( P^{(0)}_{QQ}(n) - P^{(0)}_{GG}(n) - b \right)
  \left( P^{(1)}_{QG}(n) - P^{(1)}_{QG}(n, \overline{\rm MS}) \right) + {} \nonumber\\
  +\, & P^{(0)}_{QG}(n) \left( P^{(0)}_{QQ}(n) - P^{(0)}_{GG} (n) + b \right)
  \left(P^{(1)}_{GQ}(n) - P^{(1)}_{GQ}(n, \overline{\rm MS}) \right) - {} \nonumber\\
  -\, & 2 P^{(0)}_{QG}(n) P^{(0)}_{GQ}(n) \left( P^{(1)}_{QQ} (n)
  - P^{(1)}_{GG}(n) -  P^{(1)}_{QQ} (n, \overline{\rm MS}) + P^{(1)}_{GG}(n,
  \overline{\rm MS}) \right) = 0 \label{conpracapp}
\end{align}
for $n$ for which
\begin{equation}
  b^2 - \left( P^{(0)}_{QQ}(n) - P^{(0)}_{GG}(n) \right)^2 -
  4P^{(0)}_{QG} (n) P^{(0)}_{GQ} (n) = 0 \qquad\text{and}\qquad
  \re n > 1. \label{consforn}
\end{equation}
In the case of five massless quark flavours, the approximate values of $n$
satisfying the preceding condition (\ref{consforn}) are $1.9001$ and $3.1798$.
The condition (\ref{conpracapp}) depends only on singlet splitting functions,
and therefore it does not put any constraints on the choice of non-singlet
NLO splitting functions. Hence, the choice of non-singlet NLO splitting
functions cannot cause any unexpected restrictions of practical applicability.
Since an approximately ZERO factorization scheme is required to be applicable
in the full range needed for QCD phenomenology, the influence of NLO splitting
functions on the evolution of parton distribution functions should be
minimized on a set of NLO splitting functions that satisfy the condition
of practical applicability (\ref{conpracapp}).

The difference between $\mathbf{D}_0(x, M, {\rm FS}, M_{\rm S})$ and
$\mathbf{D}(x, M, {\rm FS})$ was minimized on the set of factorization
schemes in which the non-singlet NLO splitting functions vanish and
the matrix of the singlet NLO splitting functions
$\mathbf{P}^{(1)} (x)$ is expressed as
\begin{equation}
  \mathbf{P}^{(1)}(x) = \mathbf{P}^{(1)}_0 (x) + \sum_{k=1}^{N} \lambda_k
  \mathbf{P}^{(1)}_k (x)    \label{sfsnlofceform}
\end{equation}
where the real parameters $\lambda_k$ are completely arbitrary. All matrices in
the preceding relation are square matrices of dimension 2. The set in which
an approximately ZERO factorization scheme is sought is characterized by the
functions $\mathbf{P}^{(1)}_k (x)$, $k \ge 0$ which have to be chosen such that
the singlet NLO splitting functions $\mathbf{P}^{(1)} (x)$ satisfy:
\begin{itemize}
  \item the condition of practical applicability (\ref{conpracapp}),
  \item $ \displaystyle \int_0^1 x\left( P^{(1)}_{QQ}(x) + P^{(1)}_{GQ}(x) \right) {\rm d}x = 0,
    \quad \int_0^1 x\left( P^{(1)}_{QG}(x) + P^{(1)}_{GG}(x) \right) {\rm d}x = 0.$
\end{itemize}
Both conditions must be satisfied independently of the values of the parameters
$\lambda_k$. The second condition, which is satisfied trivially in the ZERO factorization
scheme, ensures that the sum of momentum fractions of all partons is independent of
the factorization scale. The difference between $\mathbf{D}_0(x, M, {\rm FS}, M_{\rm S})$
and $\mathbf{D}(x, M, {\rm FS})$ was minimized for several different choices of
the functions $\mathbf{P}^{(1)}_k (x)$.
\begin{figure}
  \centering
  \includegraphics[width=0.4\textwidth,angle=90]{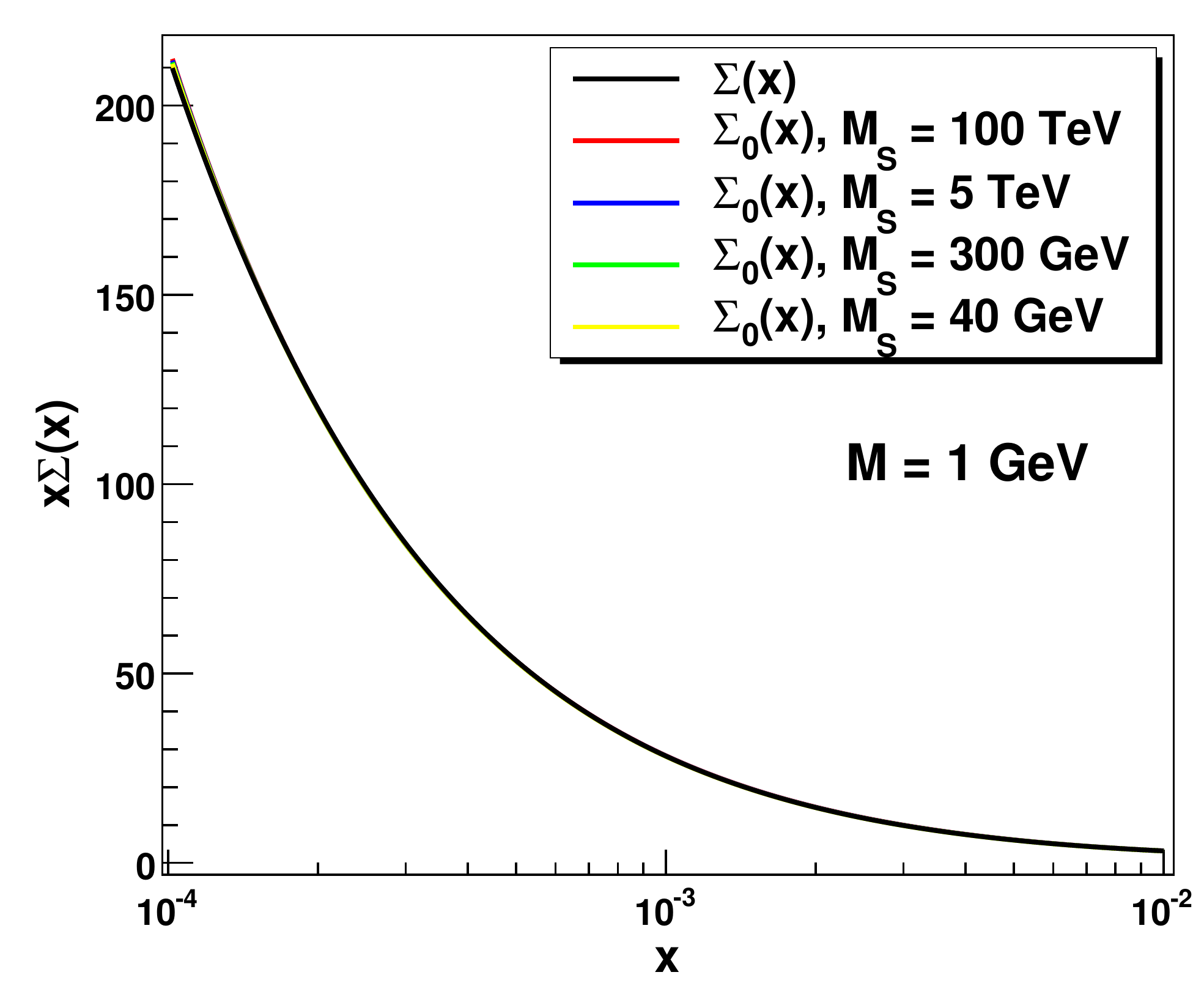}
  \includegraphics[width=0.4\textwidth,angle=90]{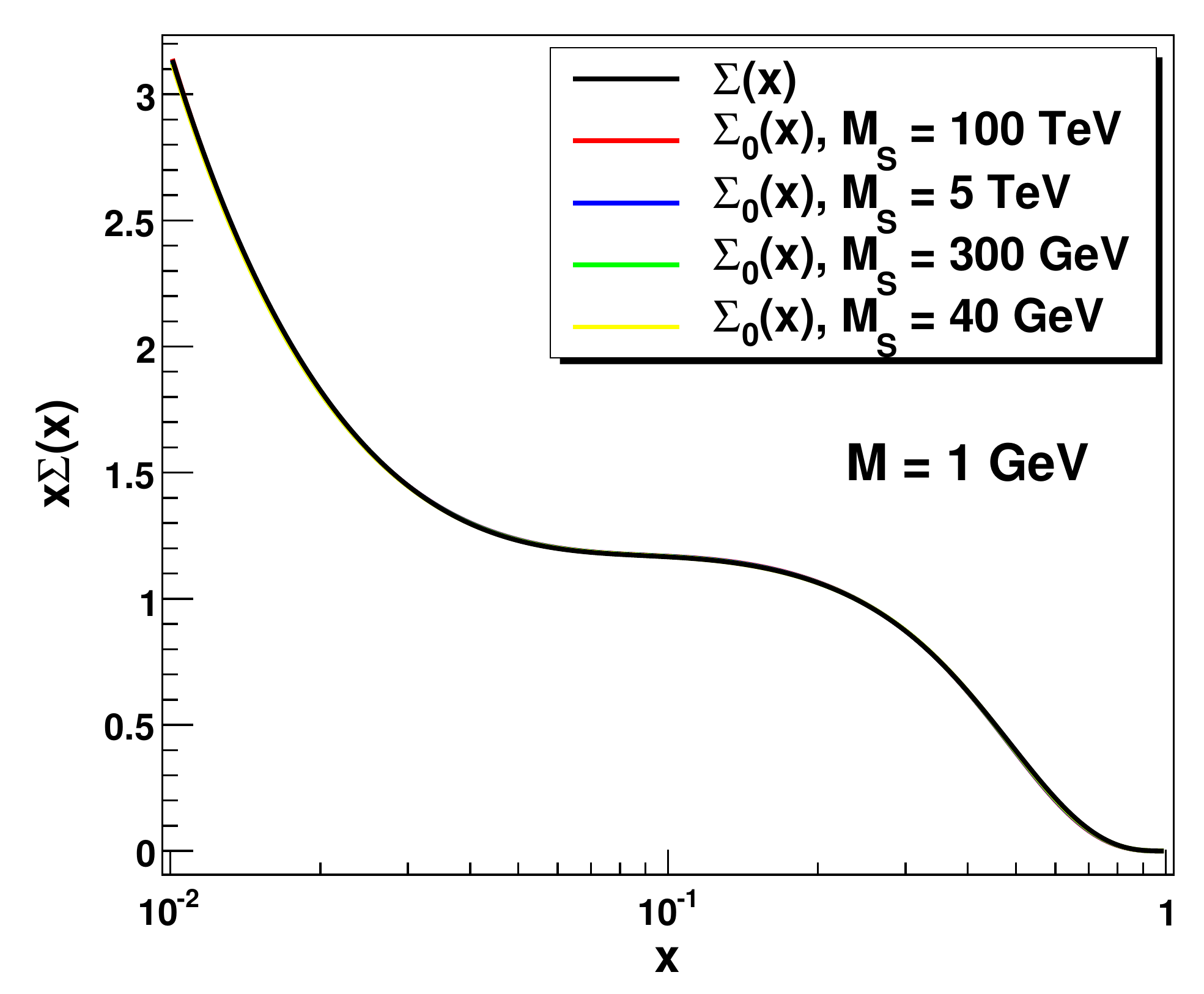}
  \includegraphics[width=0.4\textwidth,angle=90]{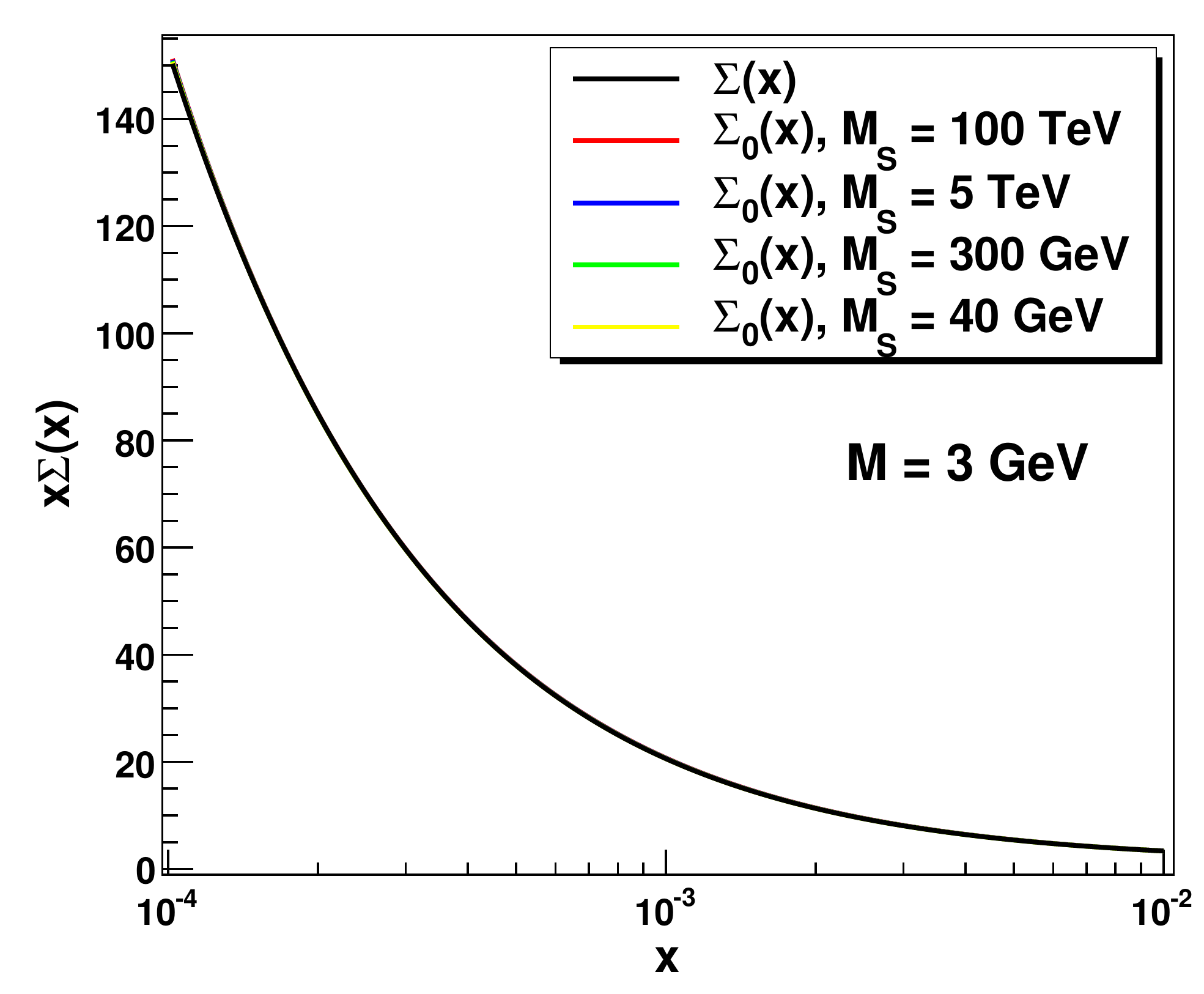}
  \includegraphics[width=0.4\textwidth,angle=90]{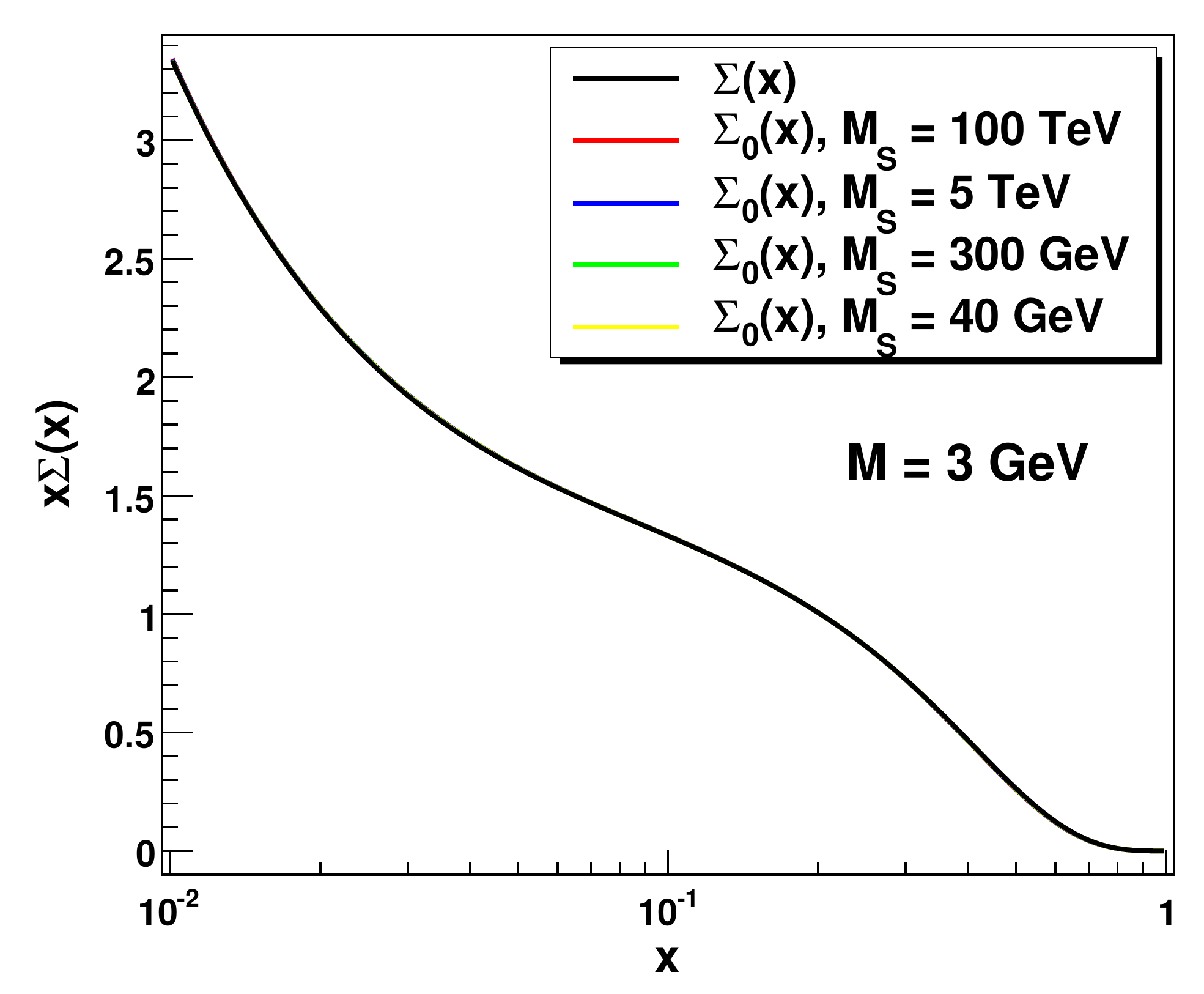}
  \includegraphics[width=0.4\textwidth,angle=90]{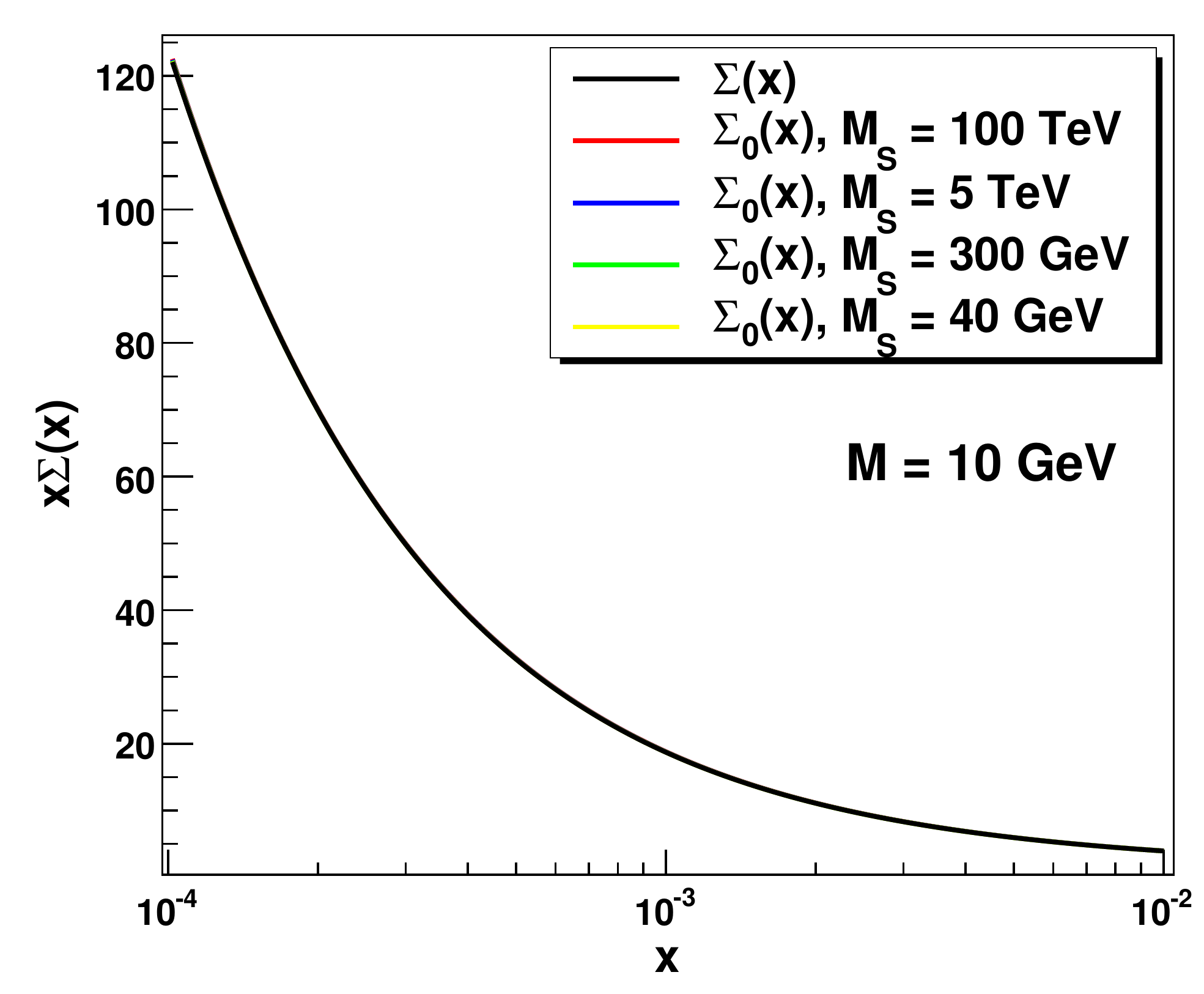}
  \includegraphics[width=0.4\textwidth,angle=90]{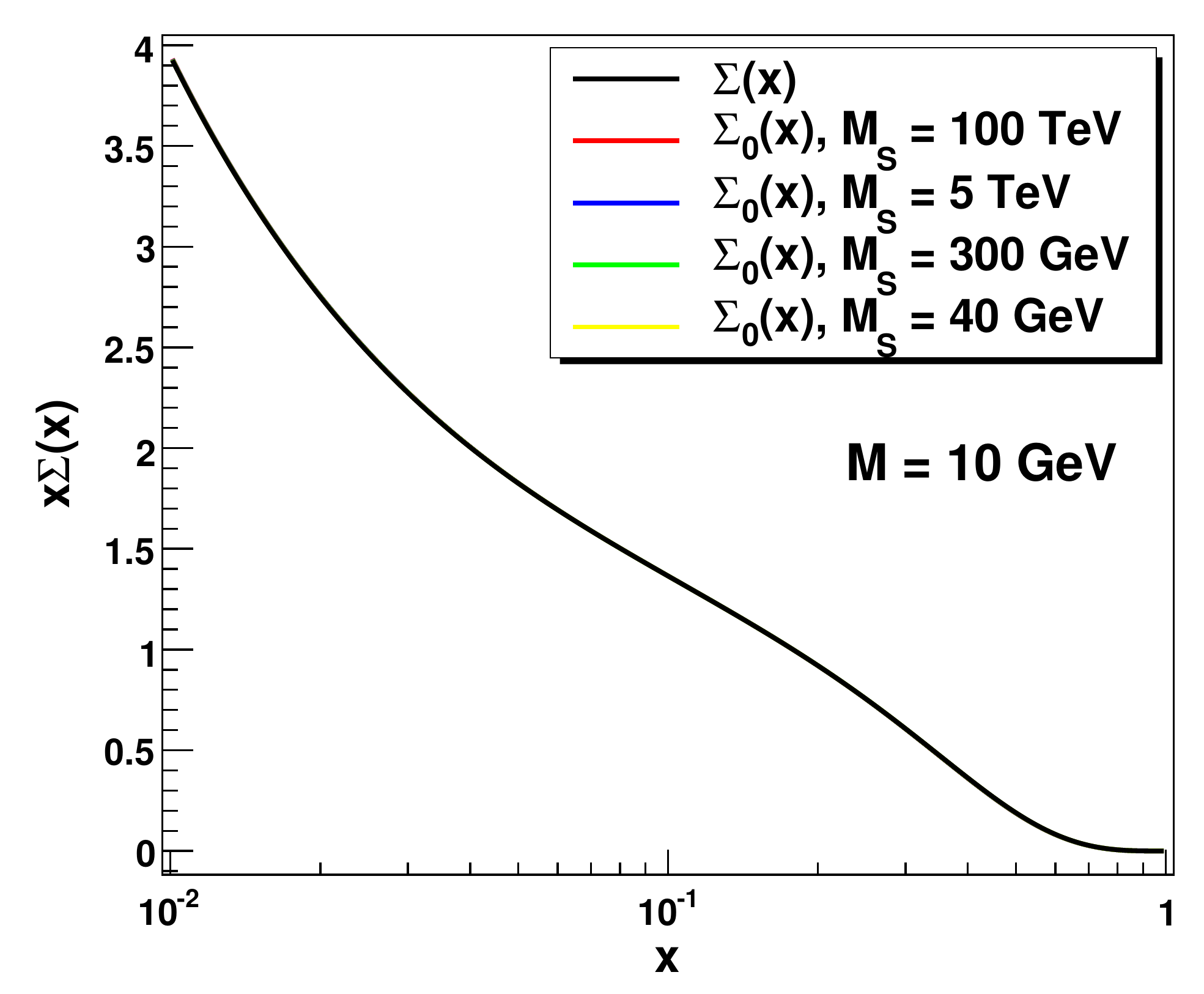}
  \caption{Comparison of the quark singlet distribution functions in the EP0 factorization scheme.
  The differences between the curves are so small that the curves overlap. Note that the ranges
  of the axes of the left graphs differ from those of the right graphs.}
  \label{figepzerosigma}
\end{figure}
\begin{figure}
  \centering
  \includegraphics[width=0.4\textwidth,angle=90]{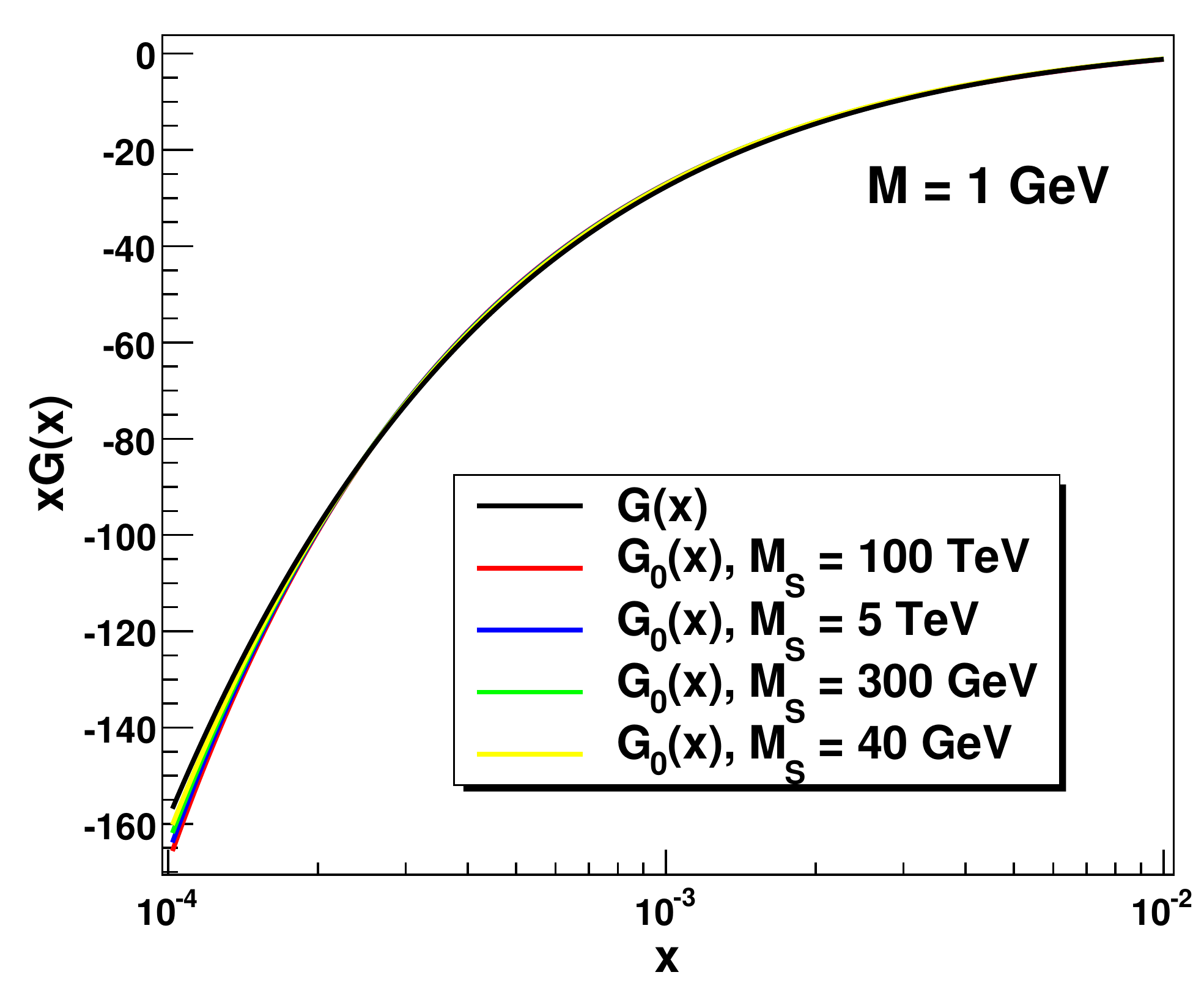}
  \includegraphics[width=0.4\textwidth,angle=90]{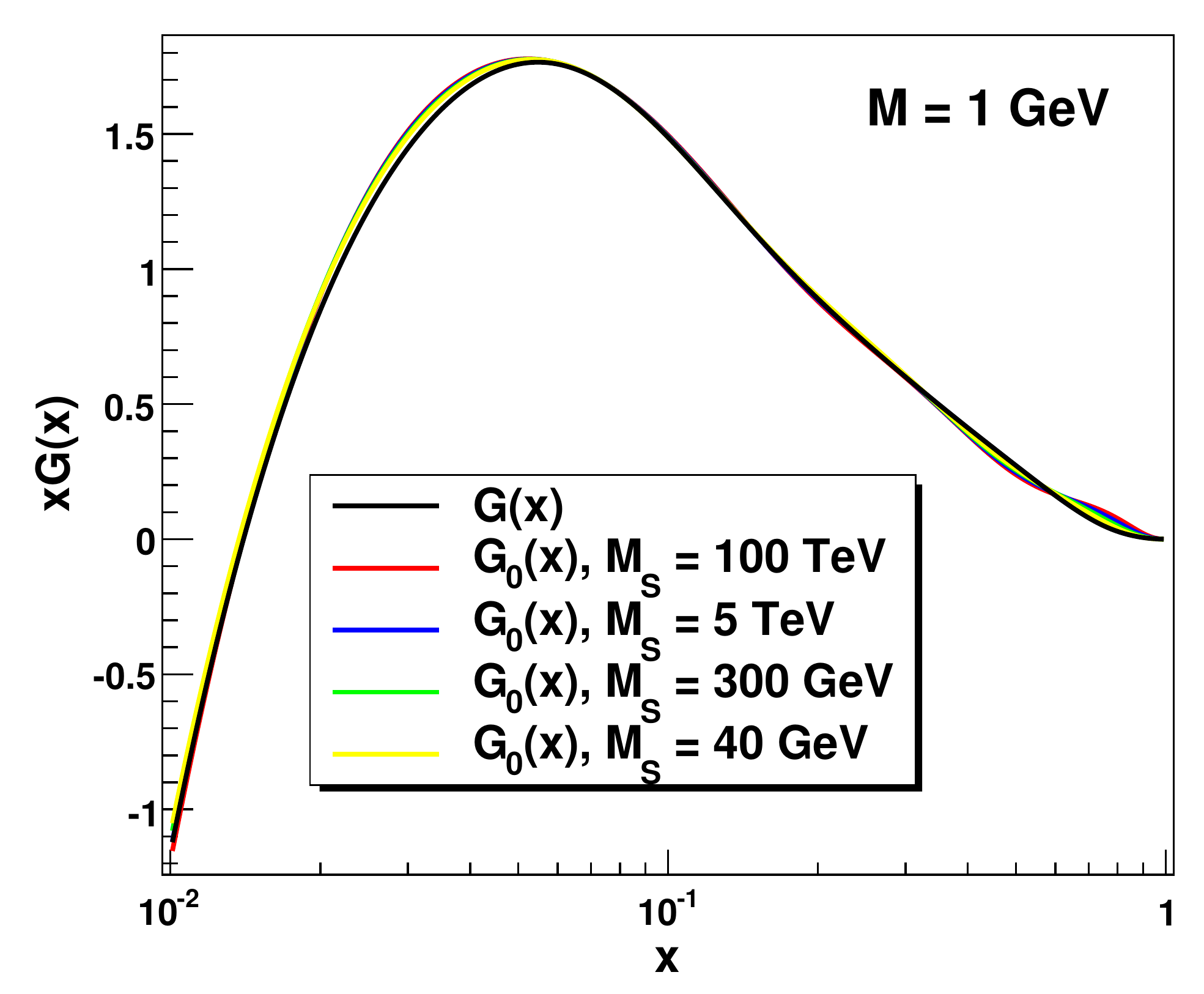}
  \includegraphics[width=0.4\textwidth,angle=90]{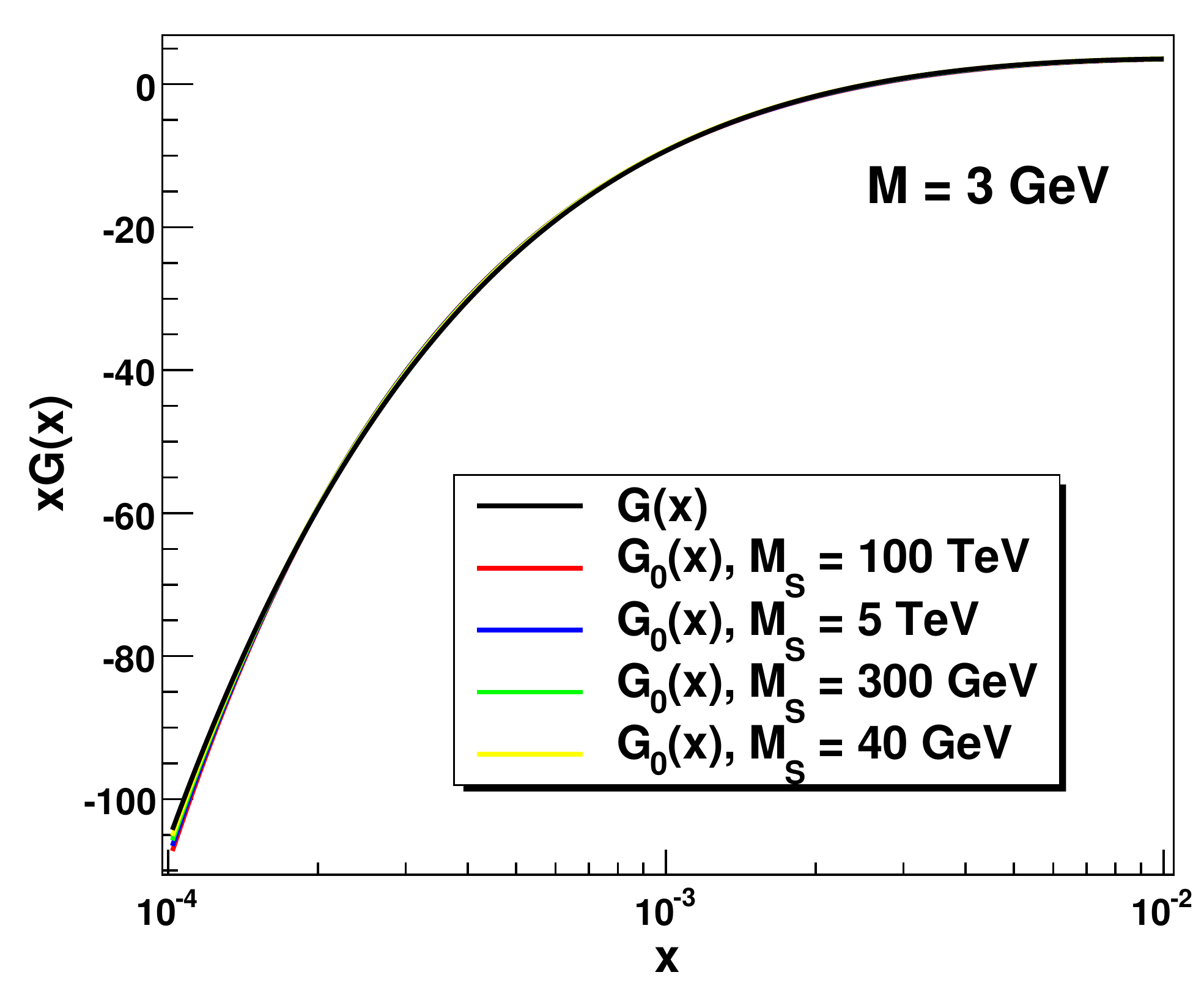}
  \includegraphics[width=0.4\textwidth,angle=90]{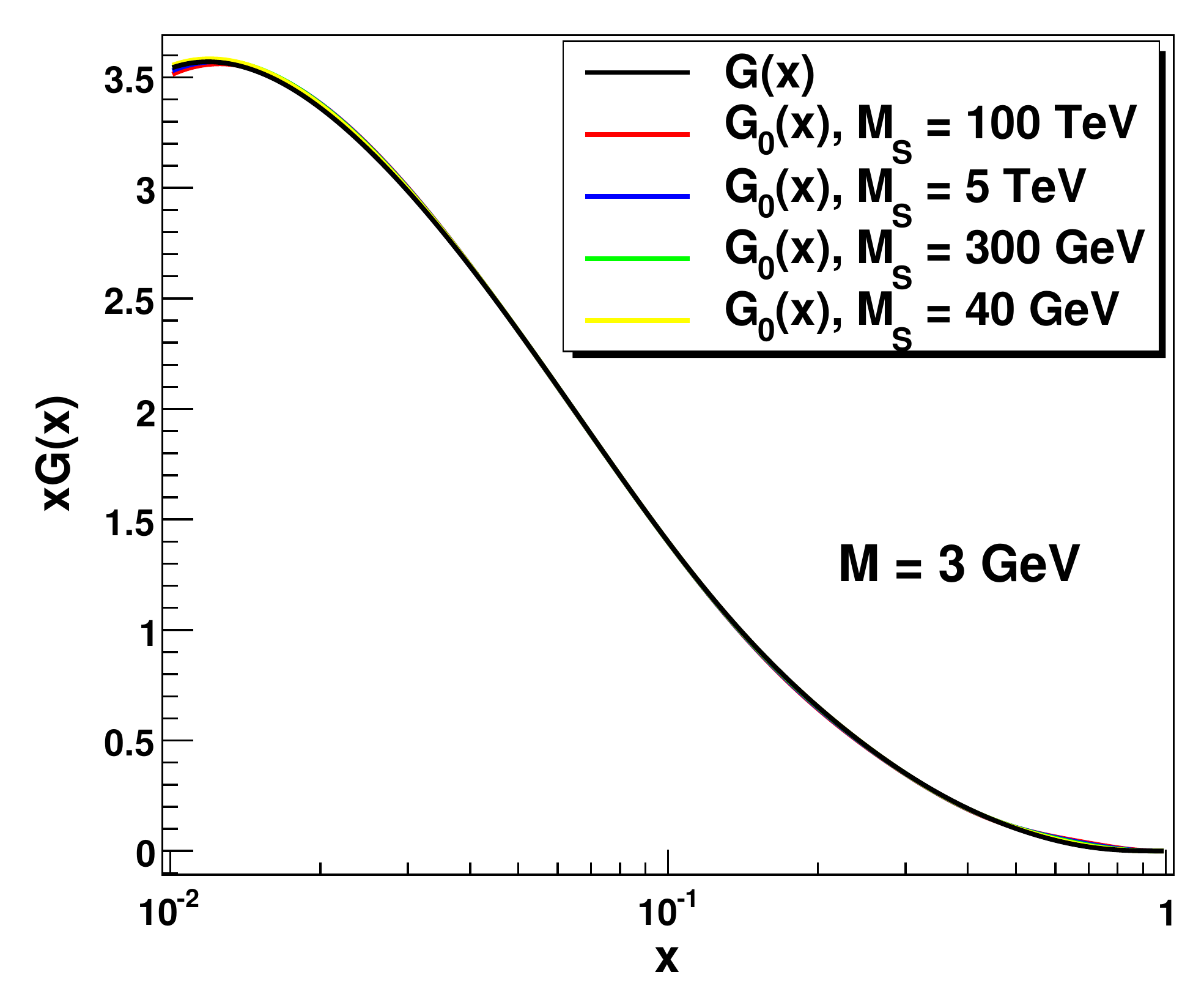}
  \includegraphics[width=0.4\textwidth,angle=90]{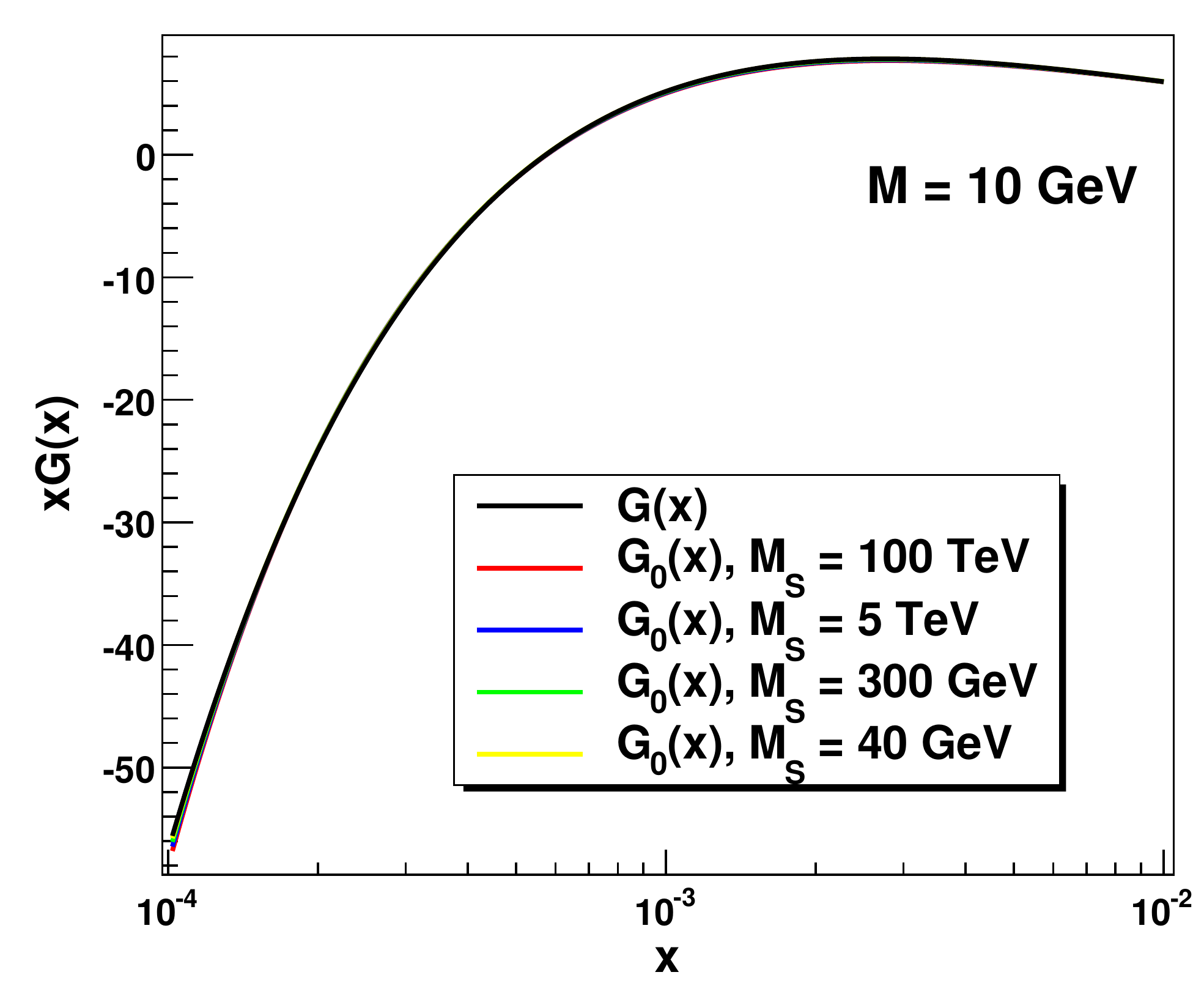}
  \includegraphics[width=0.4\textwidth,angle=90]{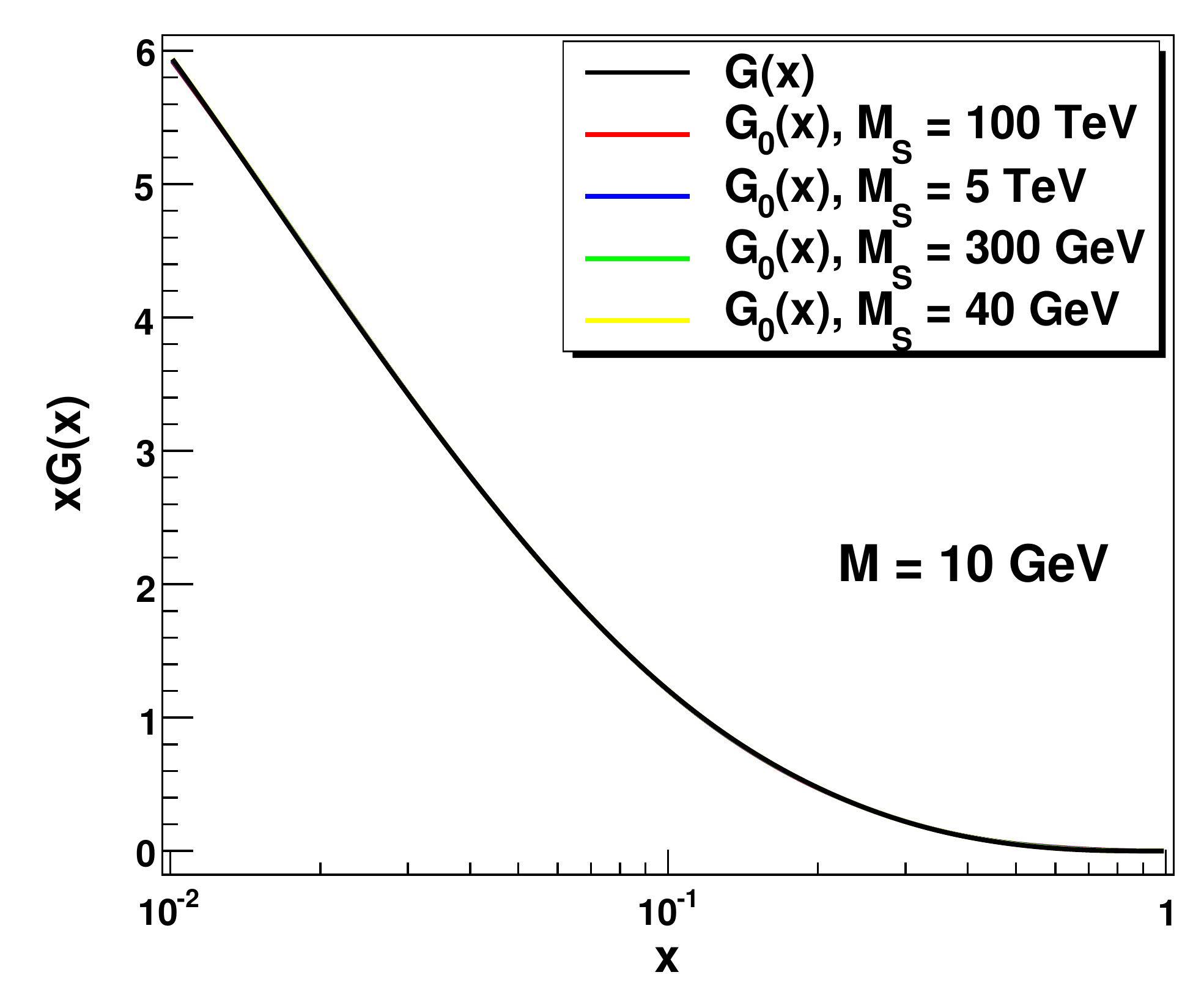}
  \caption{Comparison of the gluon distribution functions in the EP0 factorization scheme.
  Note that the ranges of the axes of the left graphs differ from those of the right graphs.}
  \label{figepzerogluon}
\end{figure}

The factorization scheme that has been found on the basis of the minimization of
the difference between $\mathbf{D}_0(x, M, {\rm FS}, M_{\rm S})$ and
$\mathbf{D}(x, M, {\rm FS})$ for five massless quark flavours will be denoted as
EP0 in the following. The comparison of the parton distribution functions
$\mathbf{D}(x, M, {\rm EP0})$ to the ``zero'' parton distribution functions
$\mathbf{D}_0 (x, M, {\rm EP0}, M_{\rm S})$ is displayed in figures \ref{figepzerosigma}
and \ref{figepzerogluon}. The non-singlet parton distribution functions are not displayed
because any ``zero'' non-singlet parton distribution function
$q_0^{\rm NS}(x, M, {\rm EP0}, M_{\rm S})$ is identical to the appropriate non-singlet
parton distribution function $q^{\rm NS}(x, M, {\rm EP0})$, which follows from the fact
that the non-singlet NLO splitting functions vanish in the EP0 factorization scheme.
The parton distribution functions are compared at small factorization
scales $M$ because the differences between $\mathbf{D}(x, M, {\rm EP0})$ and
$\mathbf{D}_0 (x, M, {\rm EP0}, M_{\rm S})$ increase with increasing distance from the
initial condition at the factorization scale $M_{\rm S}$.\footnote{Since the evolution
of initial state parton showers is backward in Monte Carlo event generators, the parton
distribution functions are compared at factorization scales $M$ smaller than
$M_{\rm S}$.} The parton distribution functions in the EP0 factorization scheme were
calculated as follows. Using formula (4.5) in \cite{kolar}, the parton distribution
functions were calculated at the factorization scale $M_{\rm T} = 10^7\, {\rm GeV}$
from the $\overline{\rm MS}$ ones that corresponded to the MSTW 2008 set \cite{mstw}.
The parton distribution functions at a general factorization scale $M$ were then
obtained by the evolution from $M_{\rm T}$ to $M$ in the EP0 factorization scheme
(the EP0 factorization scheme is defined only for five massless quark flavours,
and therefore the evolution corresponded to this number of massless quark flavours
even for factorization scales smaller than the mass of the $b$ quark). Figures
\ref{figepzerosigma} and \ref{figepzerogluon} show that the influence of the NLO
splitting functions on the evolution of the parton distribution functions corresponding
to the MSTW 2008 set \cite{mstw} is almost negligible in the EP0 factorization
scheme. From the point of view of the evolution of the parton distribution functions
of the proton, the EP0 factorization scheme is thus very close to the ZERO
factorization scheme.

\begin{figure}
  \centering
  \includegraphics[width=0.4\textwidth,angle=90]{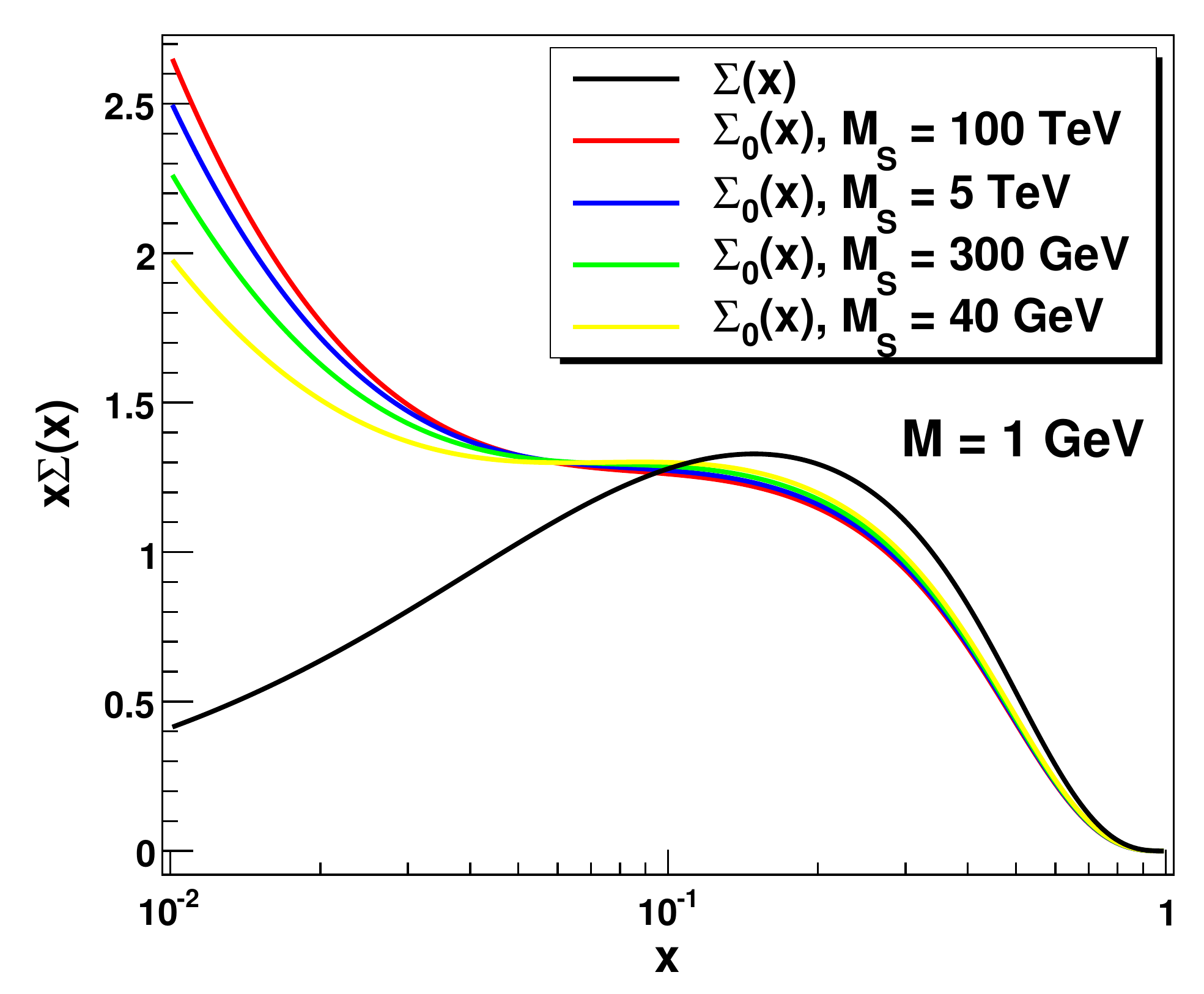}
  \includegraphics[width=0.4\textwidth,angle=90]{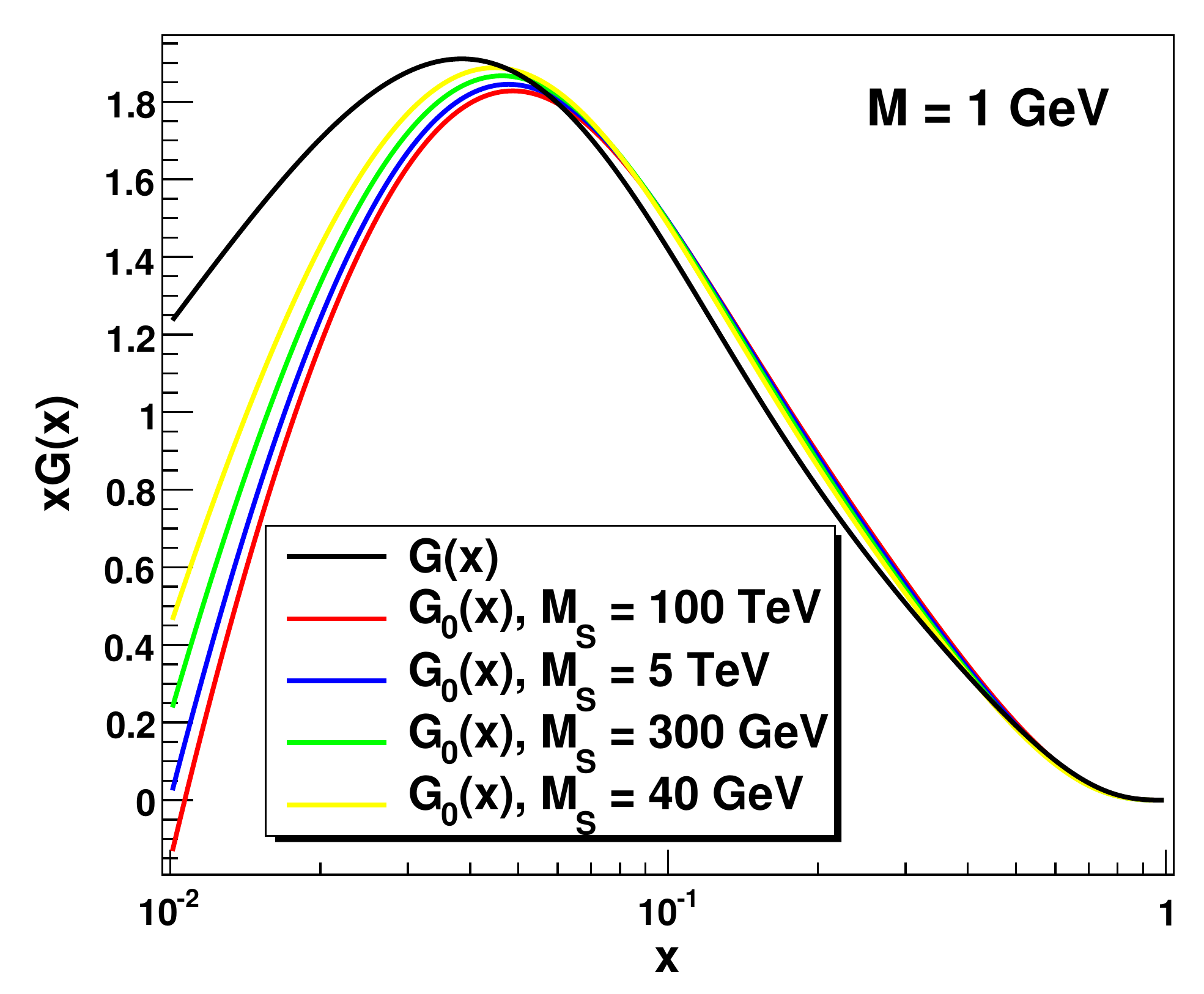}
  \includegraphics[width=0.4\textwidth,angle=90]{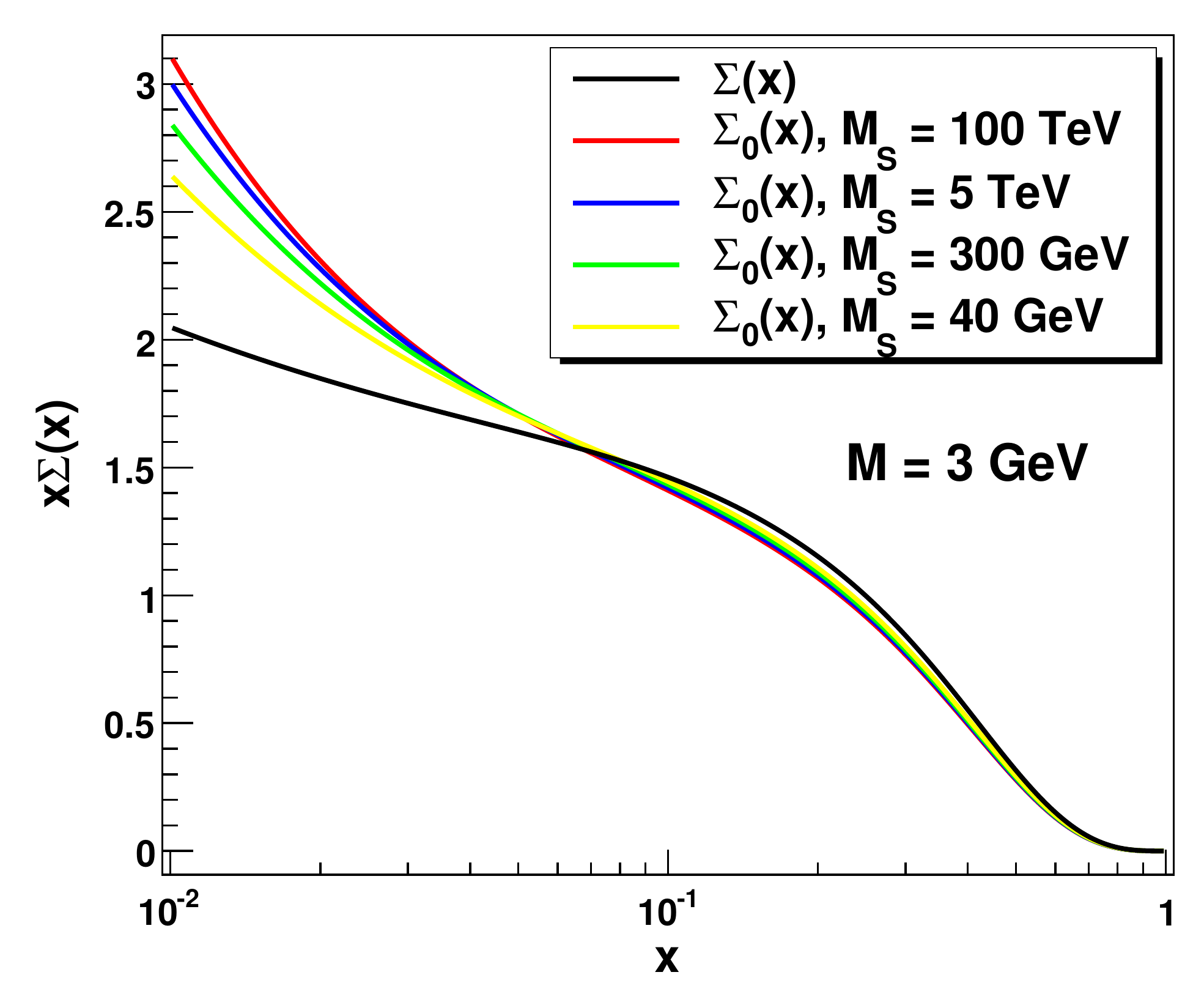}
  \includegraphics[width=0.4\textwidth,angle=90]{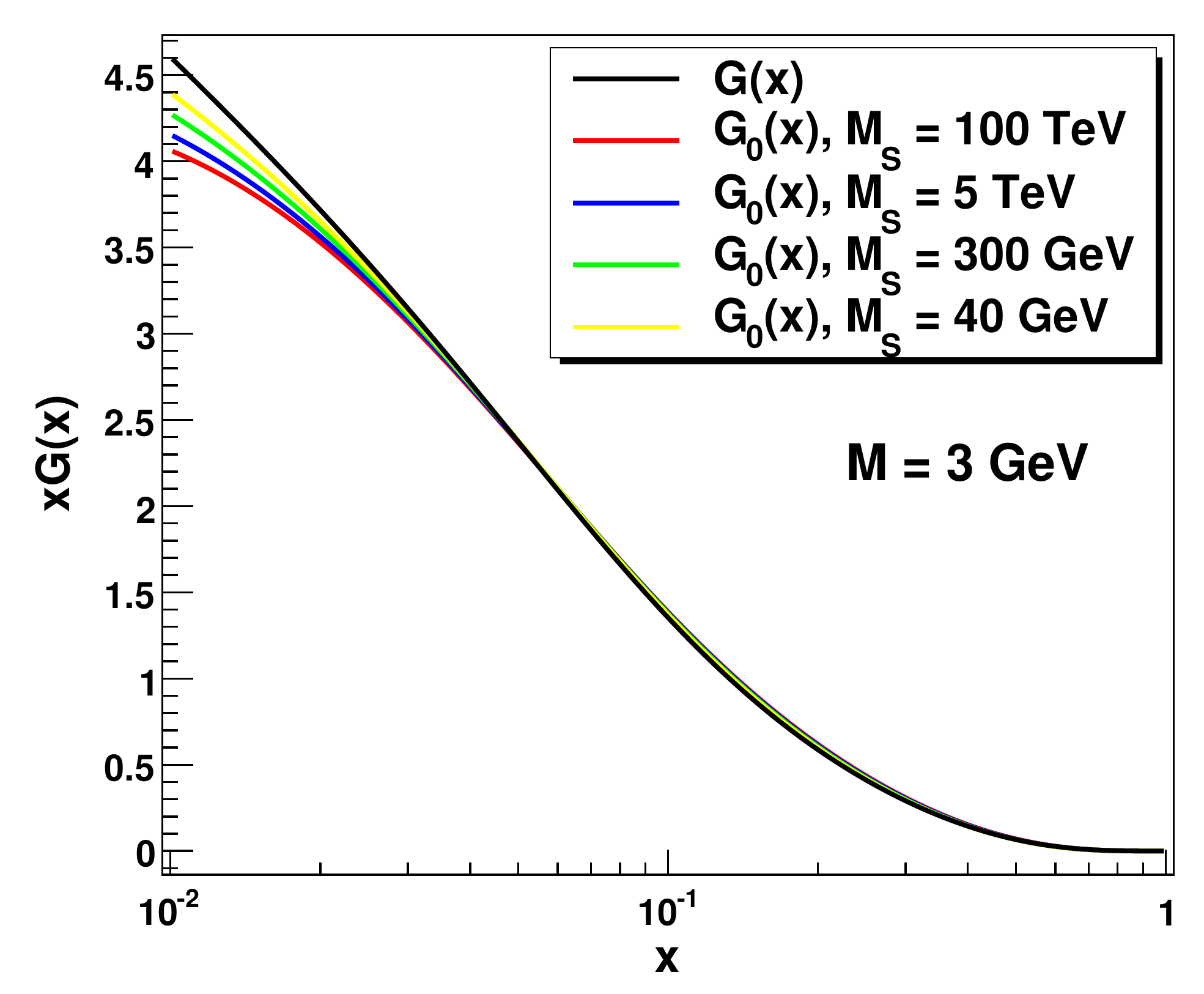}
  \includegraphics[width=0.4\textwidth,angle=90]{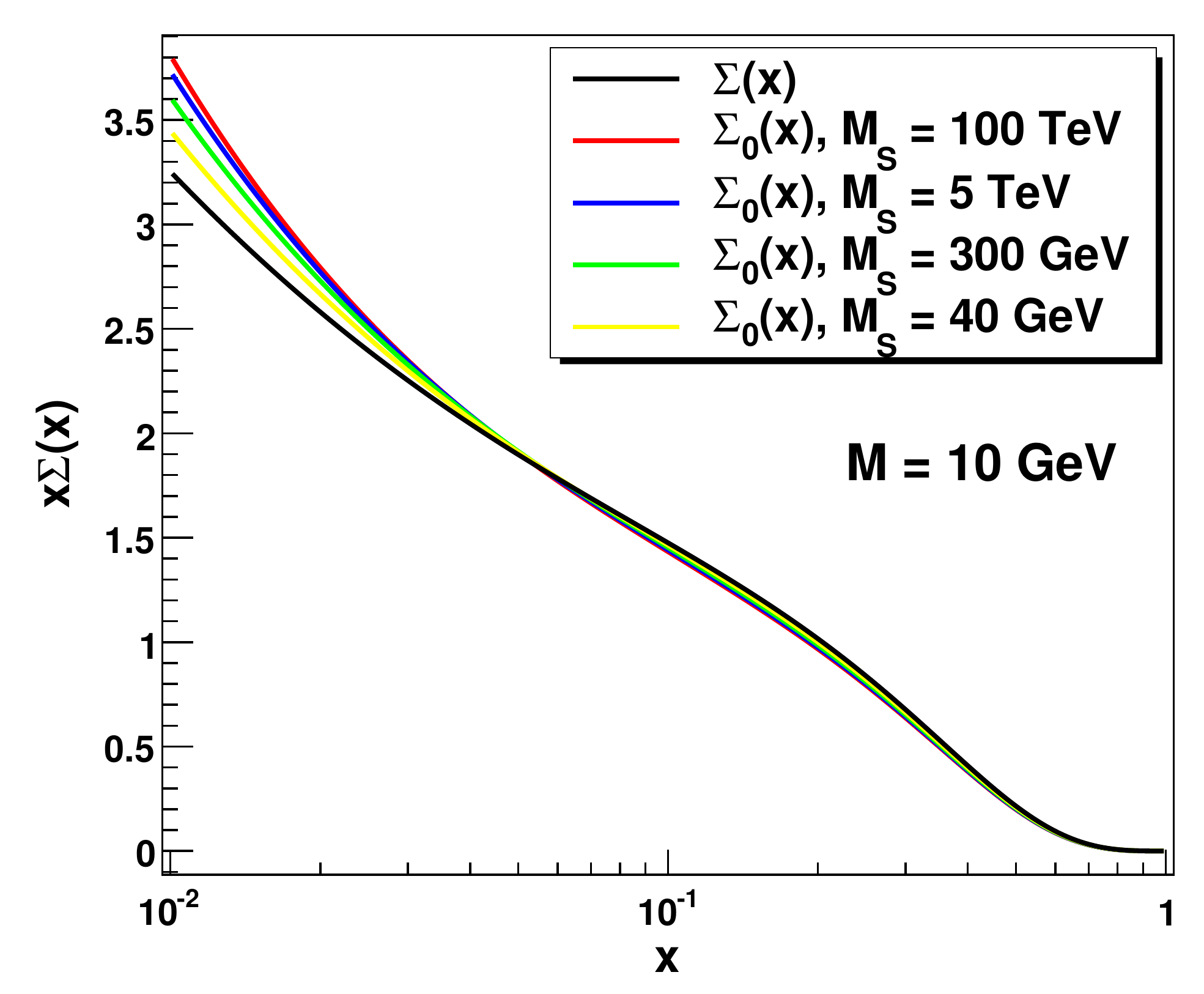}
  \includegraphics[width=0.4\textwidth,angle=90]{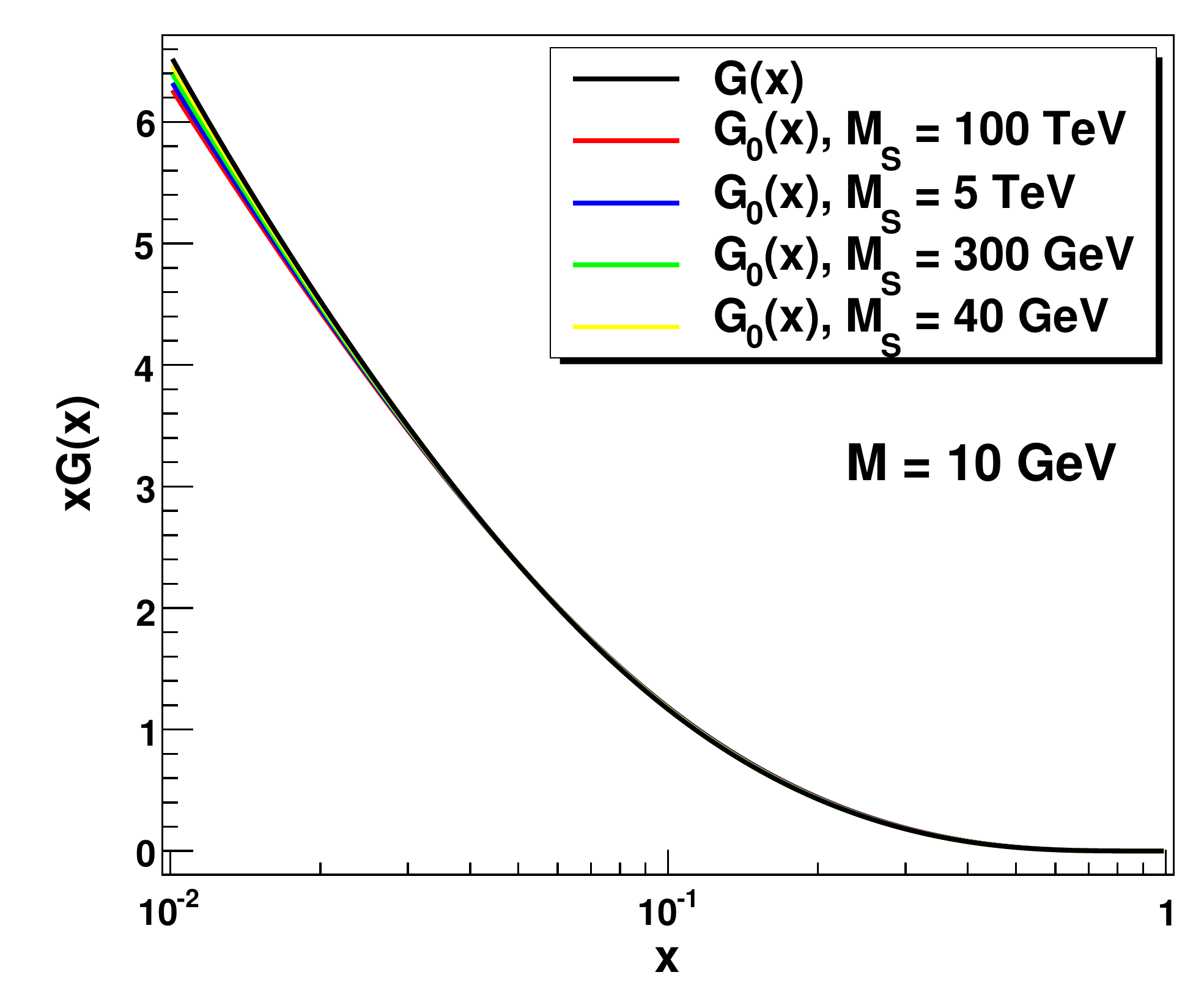}
  \caption{Comparison of the quark singlet and gluon distribution functions in
  the $\overline{\rm MS}$ factorization scheme for large $x$.}
  \label{figmsbarefzerolargex}
\end{figure}

\begin{figure}
  \centering
  \includegraphics[width=0.4\textwidth,angle=90]{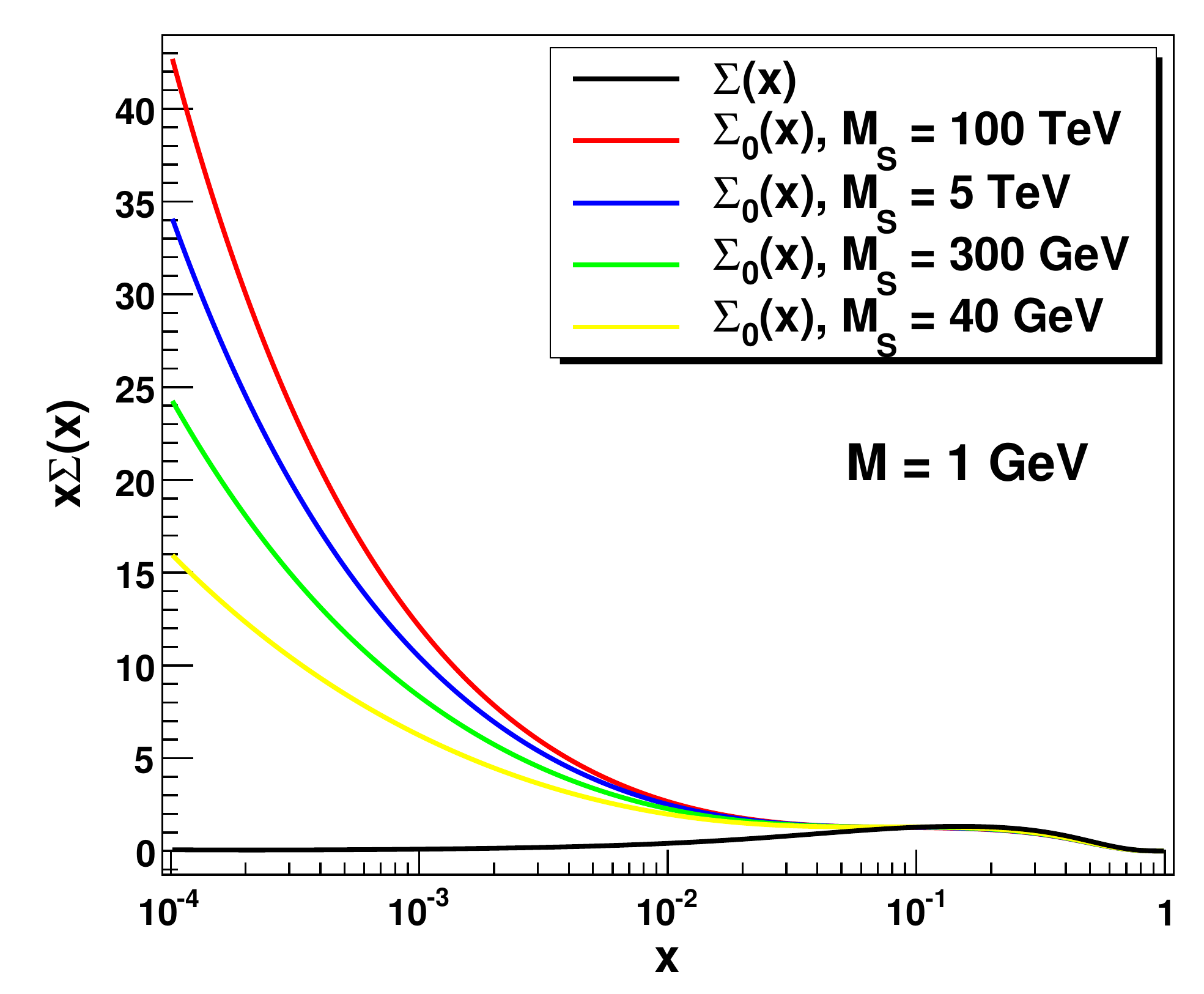}
  \includegraphics[width=0.4\textwidth,angle=90]{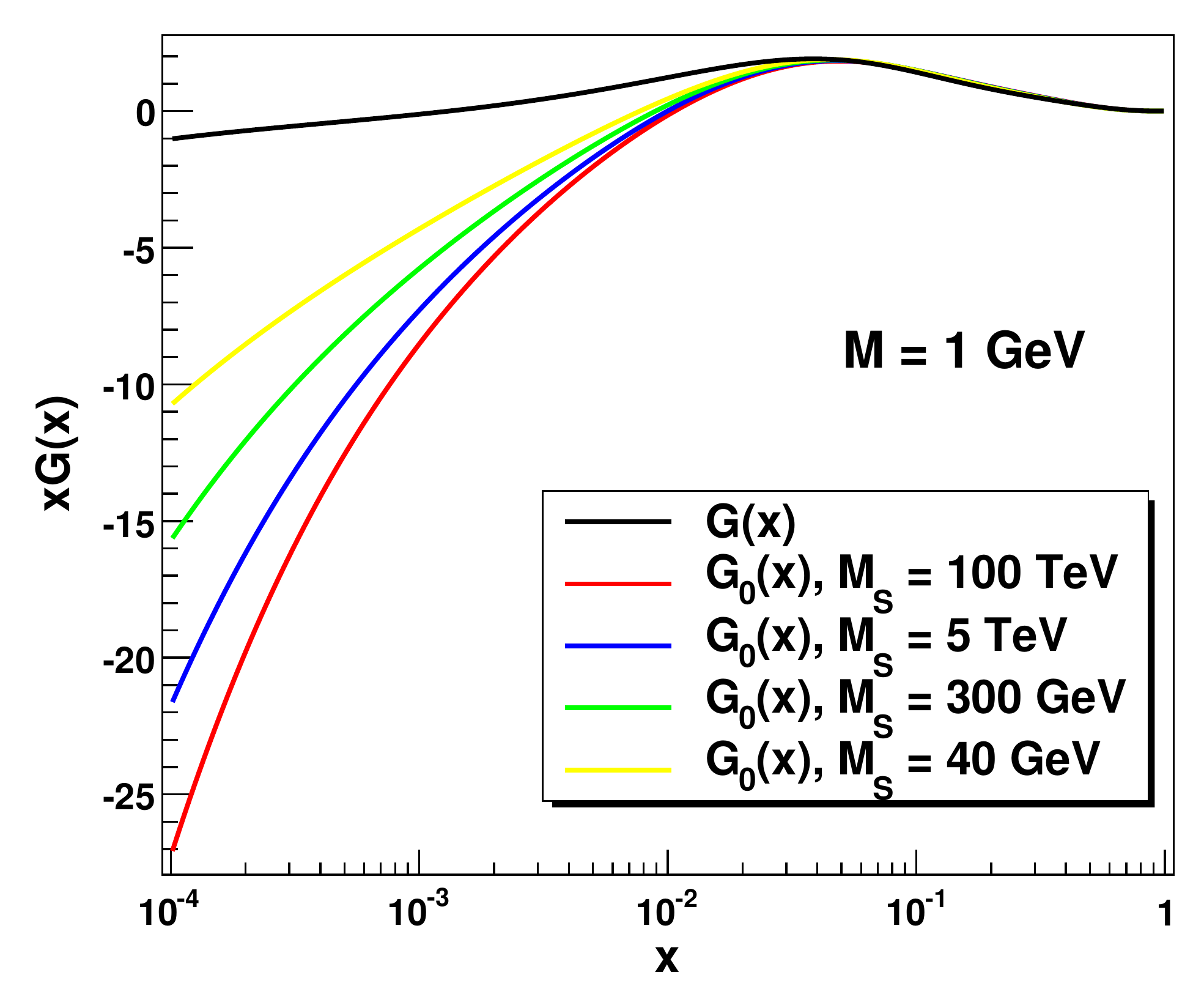}
  \includegraphics[width=0.4\textwidth,angle=90]{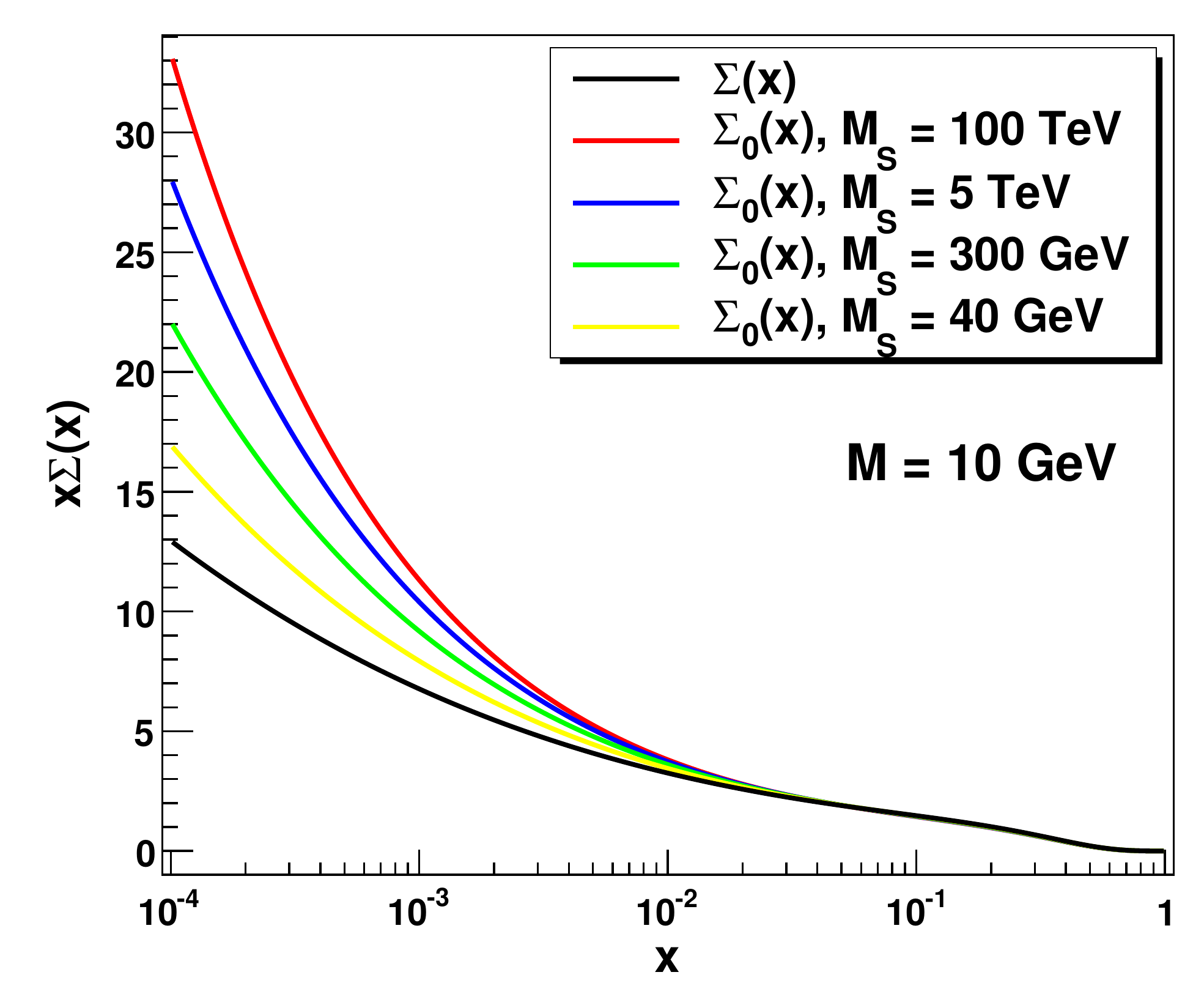}
  \includegraphics[width=0.4\textwidth,angle=90]{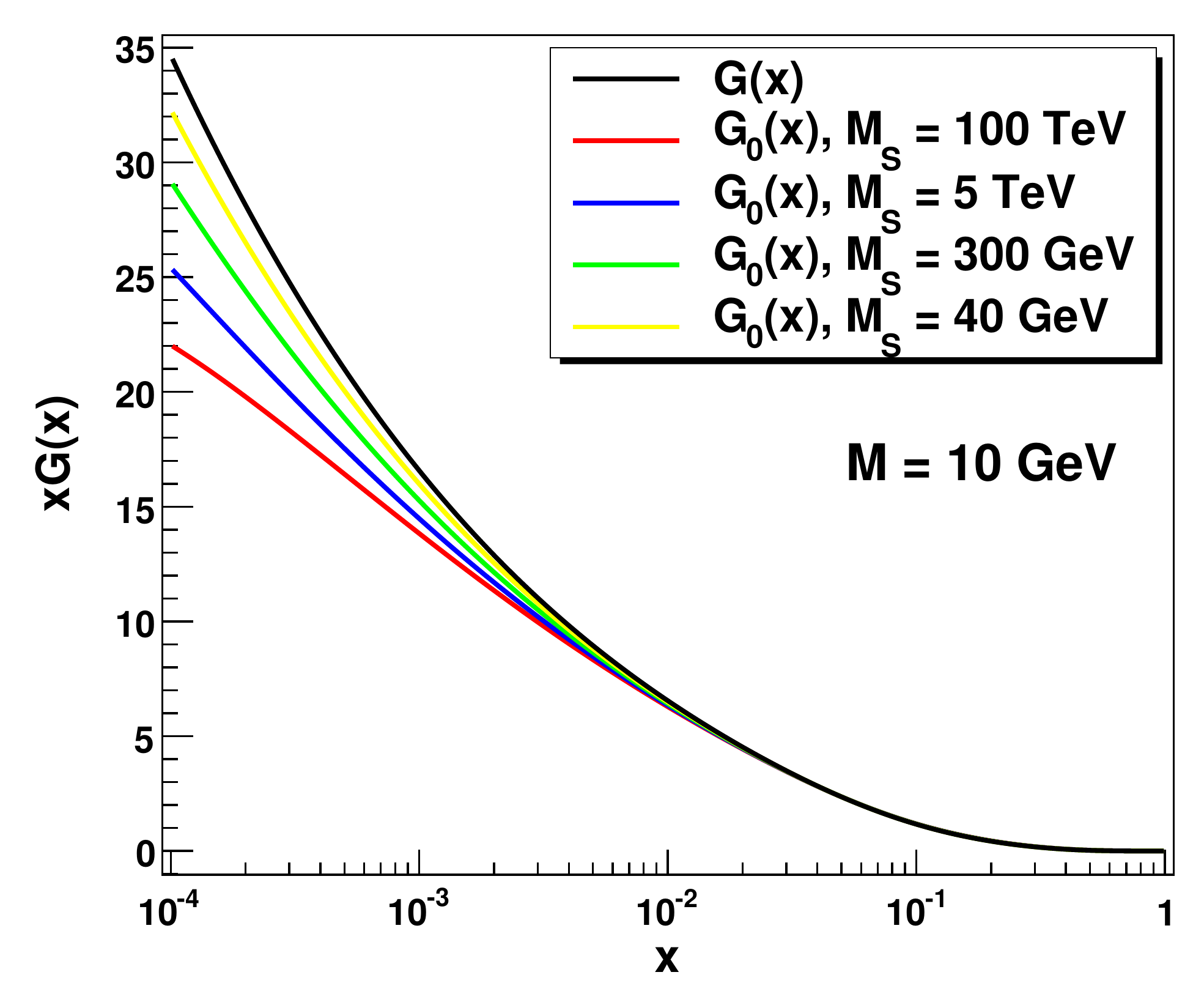}
  \includegraphics[width=0.4\textwidth,angle=90]{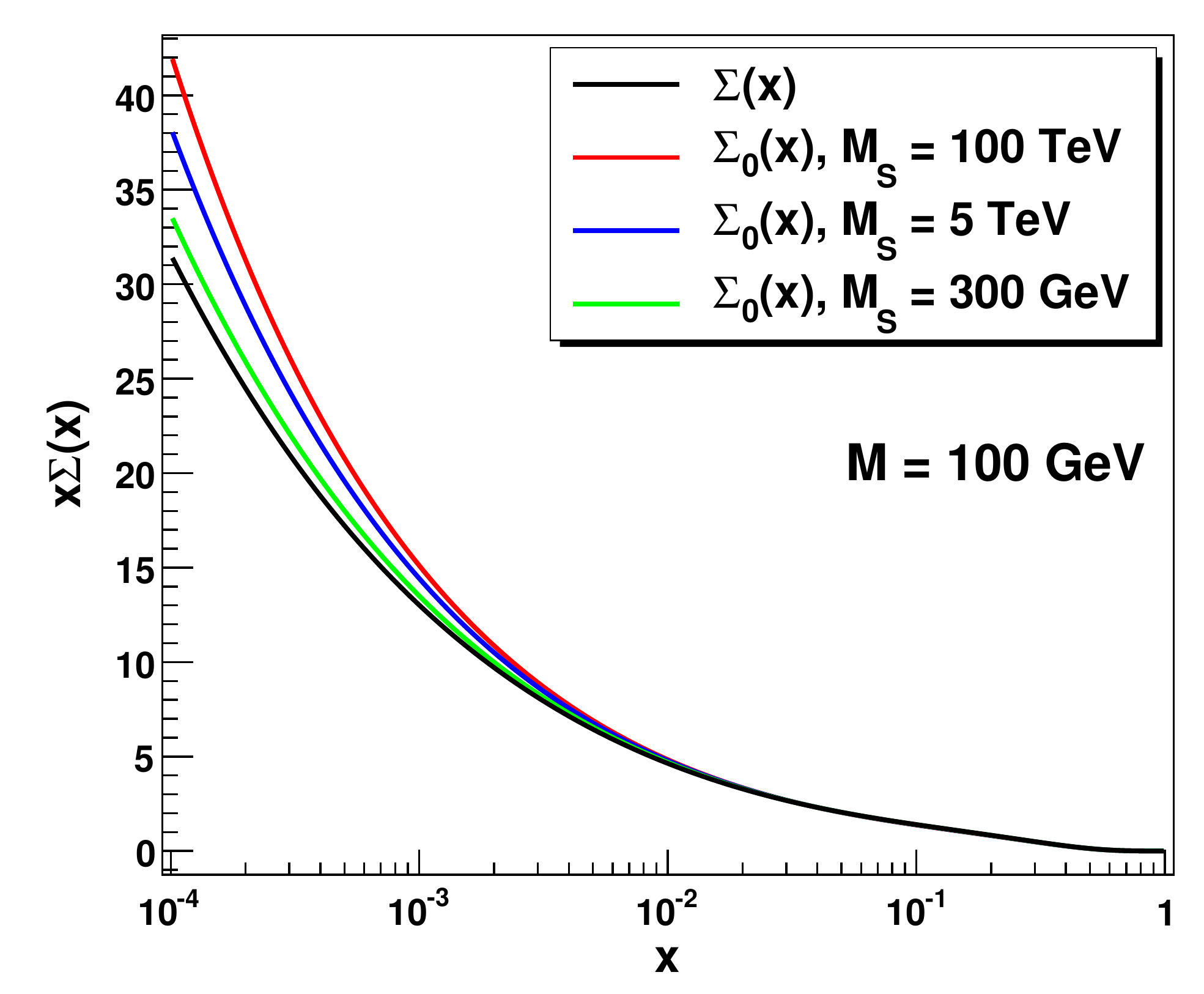}
  \includegraphics[width=0.4\textwidth,angle=90]{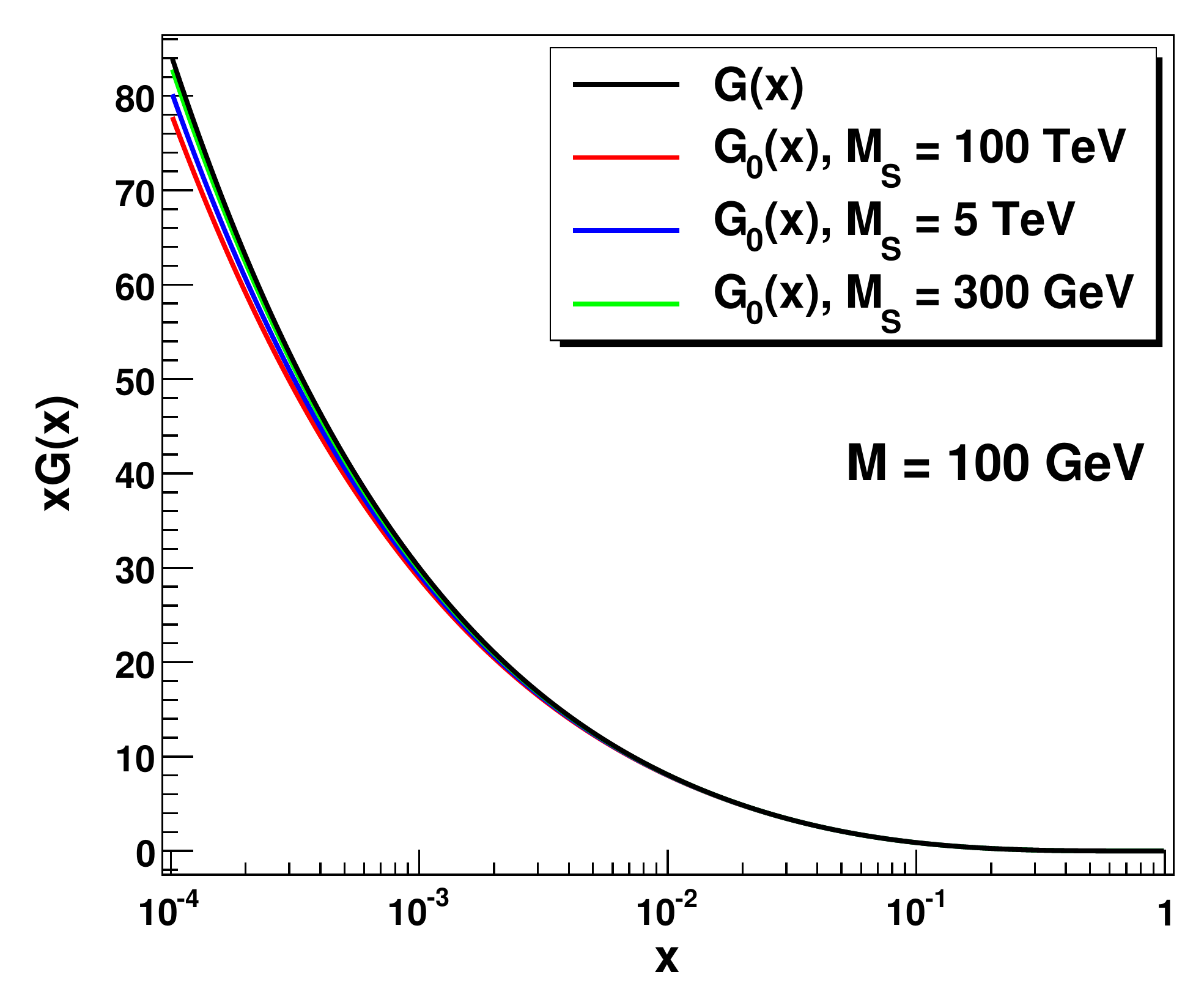}
  \caption{Comparison of the quark singlet and gluon distribution functions in
  the $\overline{\rm MS}$ factorization scheme.}
  \label{figmsbarefzero}
\end{figure}

The comparison of the parton distribution functions $\mathbf{D}(x, M, \overline{\rm MS})$
to the ``zero'' parton distribution functions $\mathbf{D}_0 (x, M, \overline{\rm MS}, M_{\rm S})$
is displayed\footnote{Similarly as in the case of the EP0 factorization scheme, only the
quark singlet and gluon distribution functions are displayed. However, contrary to the
EP0 factorization scheme, displaying the non-singlet parton distribution functions would make
sense in the case of the $\overline{\rm MS}$ factorization scheme because non-singlet
parton distribution functions $q^{\rm NS}(x, M, \overline{\rm MS})$ differ from the appropriate
``zero'' non-singlet parton distribution functions $q_0^{\rm NS}(x, M, \overline{\rm MS}, M_{\rm S})$.}
in figures \ref{figmsbarefzerolargex} and \ref{figmsbarefzero}.
The $\overline{\rm MS}$ parton distribution functions that were used for the construction of these
graphs corresponded to the MSTW 2008 set \cite{mstw} for factorization scales greater than the mass
of the $b$ quark. The parton distribution functions at factorization scales less than the mass
of the $b$ quark were then obtained by the evolution from greater factorization scales in
the $\overline{\rm MS}$ factorization scheme for five massless quark flavours (and therefore
the $\overline{\rm MS}$ parton distribution functions differ from those of the MSTW 2008 set
\cite{mstw} for factorization scales smaller than the mass of the $b$ quark). Figures
\ref{figmsbarefzerolargex} and \ref{figmsbarefzero} show that the effect of the NLO splitting
functions on the evolution of the parton distribution functions cannot be neglected in
the $\overline{\rm MS}$ factorization scheme, especially in the low $x$ region. The influence
of the NLO splitting functions on the evolution of the parton distribution functions also
manifests itself by the dependence of the ``zero'' parton distribution functions
$\mathbf{D}_0 (x, M, {\rm FS}, M_{\rm S})$ on the ``starting'' factorization scale $M_{\rm S}$.
Comparing figures \ref{figepzerosigma}--\ref{figmsbarefzero}, one sees that the ``zero''
parton distribution functions $\mathbf{D}_0 (x, M, {\rm FS}, M_{\rm S})$ depend on $M_{\rm S}$
significantly more in the case of the $\overline{\rm MS}$ factorization scheme than in the case
of the EP0 factorization scheme, for which the dependence is almost negligible. From the point
of view of the evolution of the parton distribution functions of the proton, the EP0 factorization
scheme is significantly closer to the ZERO factorization scheme than the $\overline{\rm MS}$ one.

\begin{figure}
  \centering
  \includegraphics[width=0.4\textwidth,angle=90]{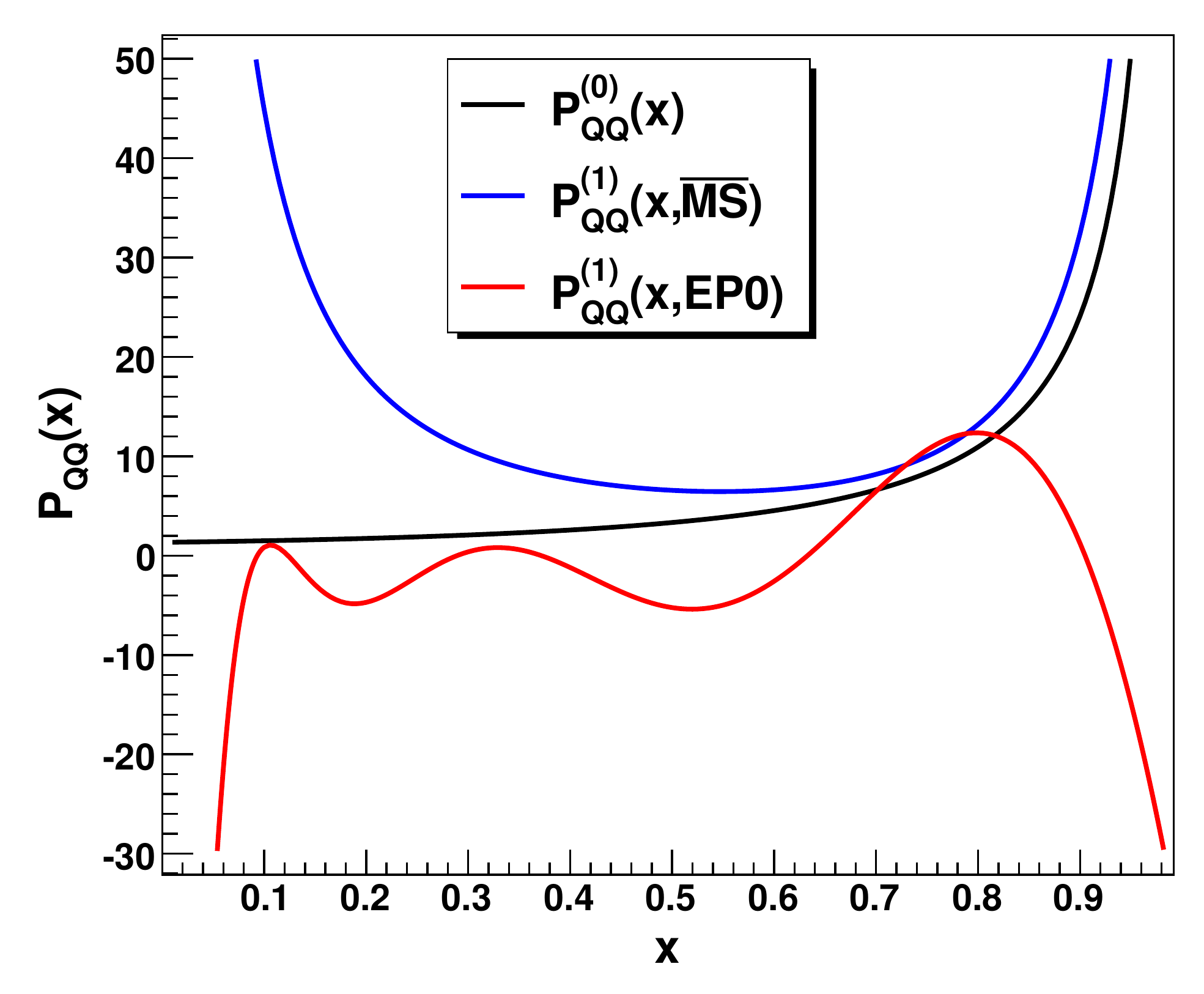}
  \includegraphics[width=0.4\textwidth,angle=90]{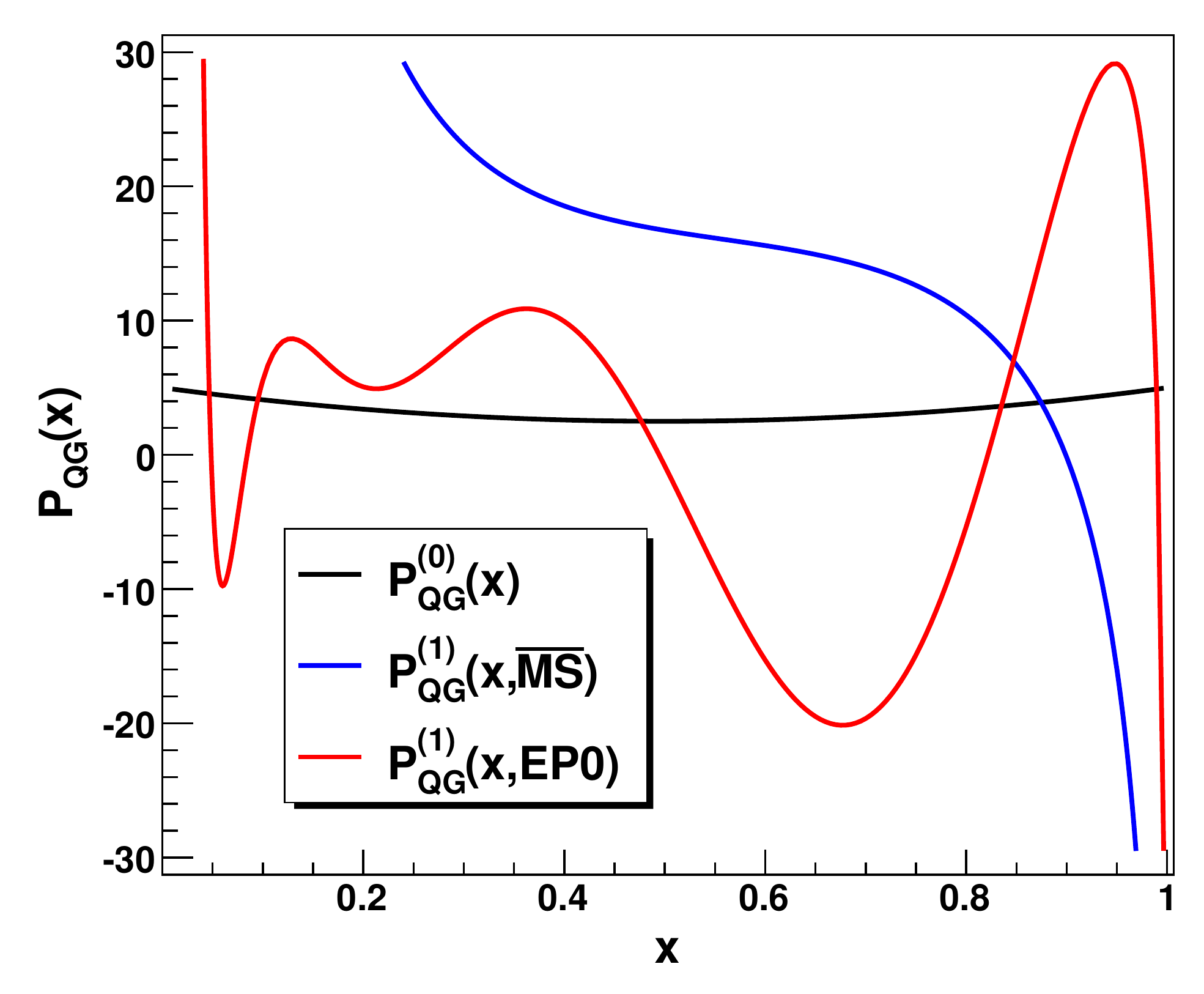}
  \includegraphics[width=0.4\textwidth,angle=90]{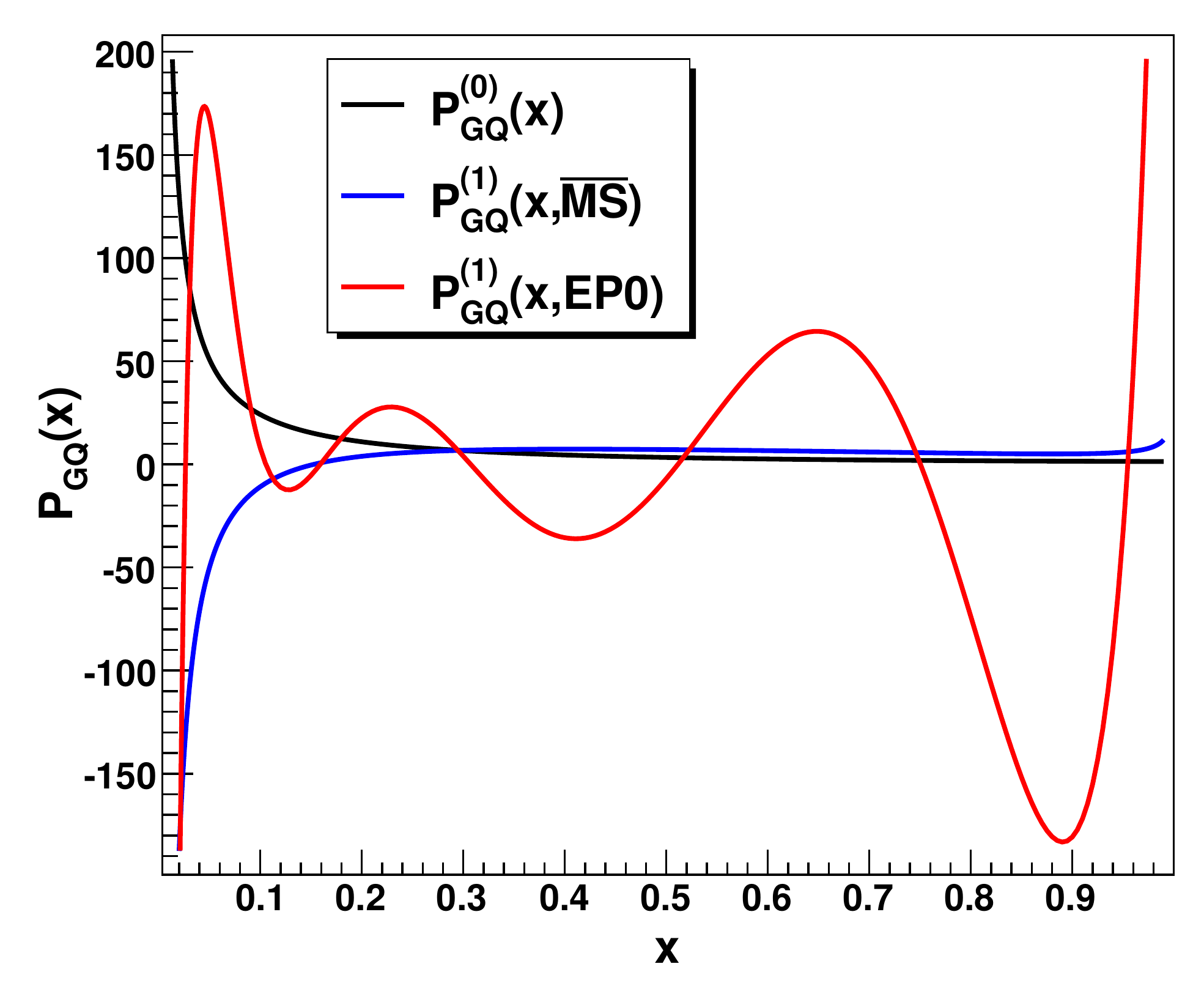}
  \includegraphics[width=0.4\textwidth,angle=90]{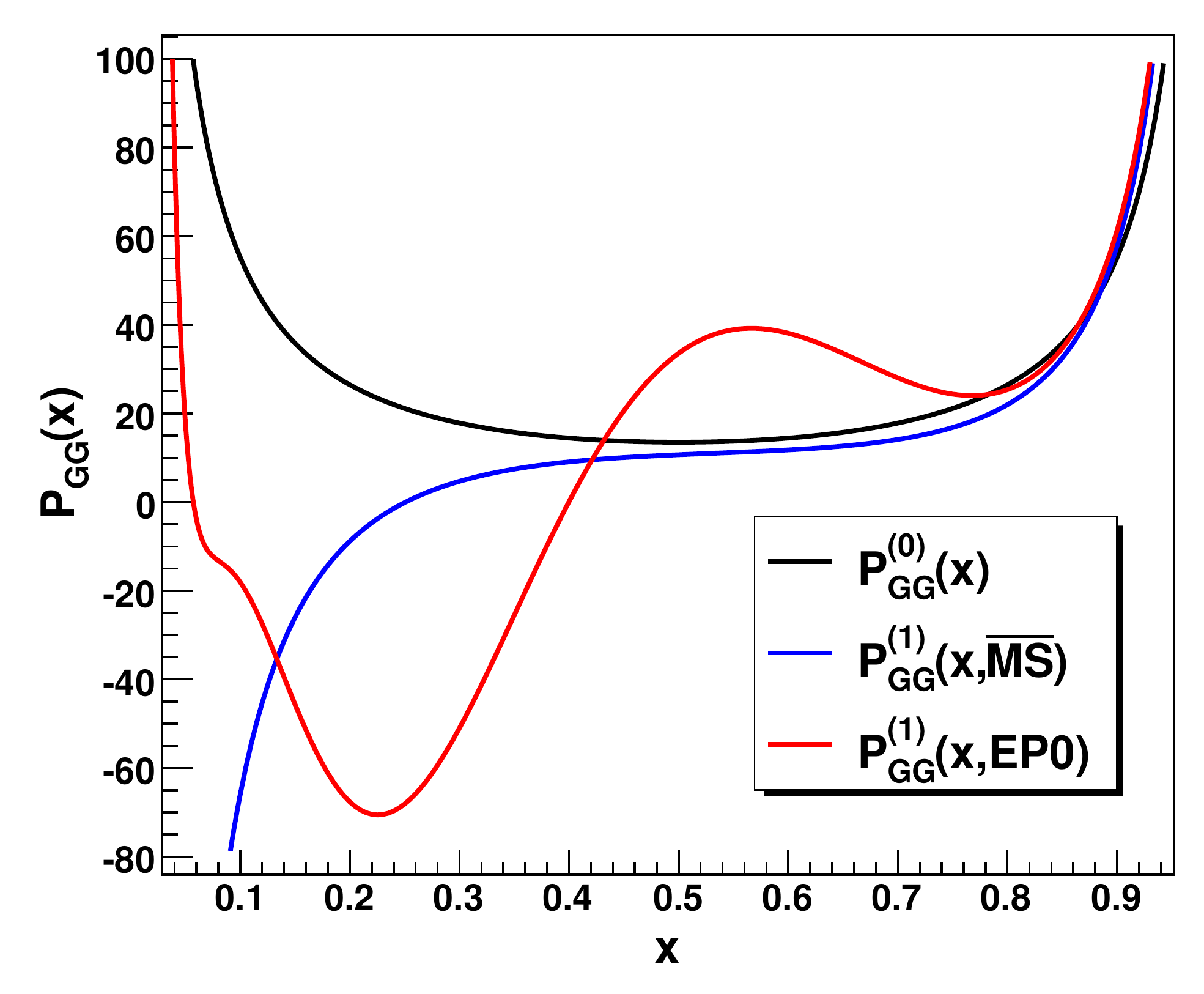}
  \caption{Comparison of singlet LO and NLO splitting functions for five massless quark flavours.
  The singlet NLO splitting functions are plotted for the $\overline{\rm MS}$ and EP0
  factorization scheme.}
  \label{figsplitfce}
\end{figure}

\begin{figure}
  \centering
  \includegraphics[width=0.4\textwidth,angle=90]{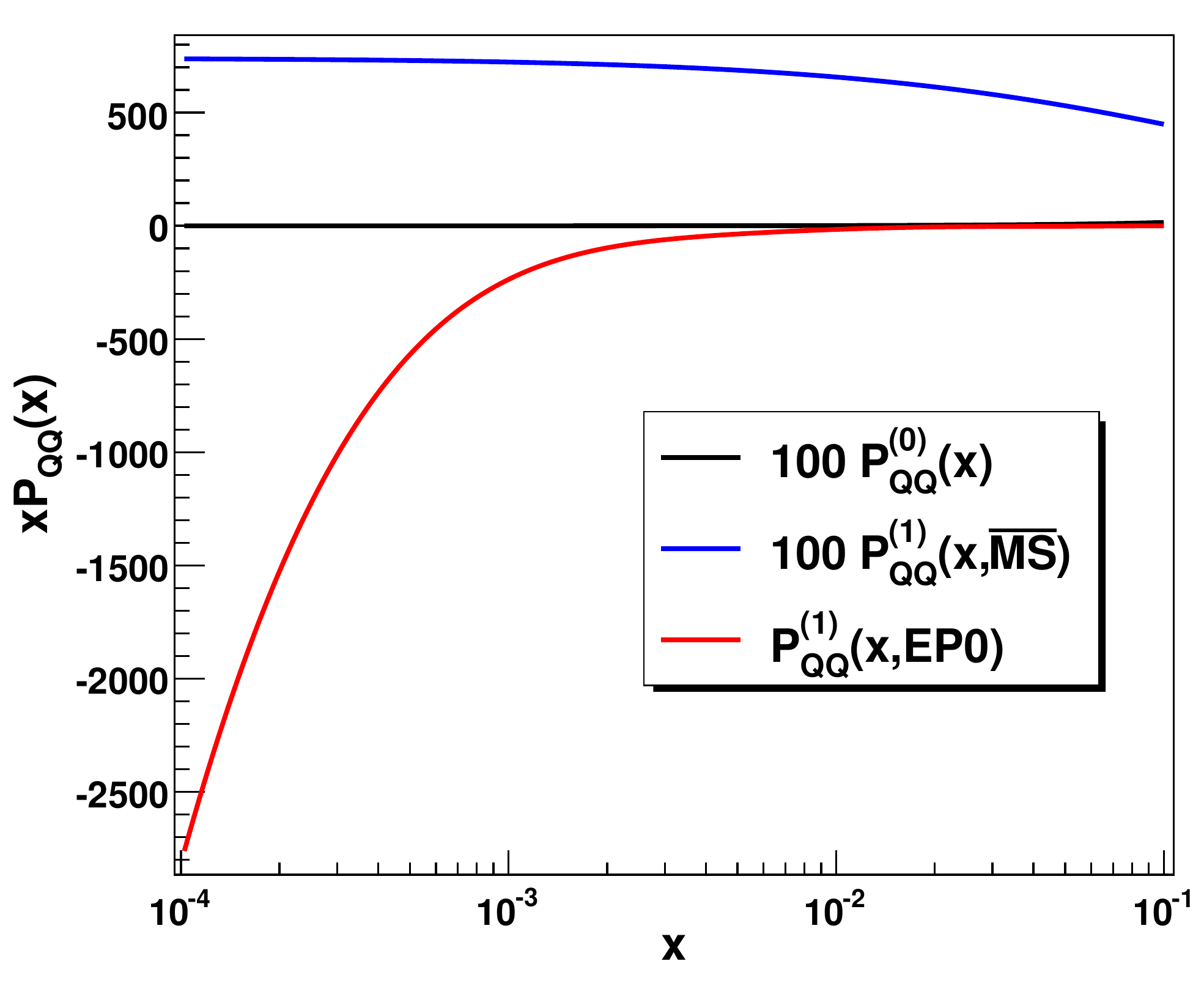}
  \includegraphics[width=0.4\textwidth,angle=90]{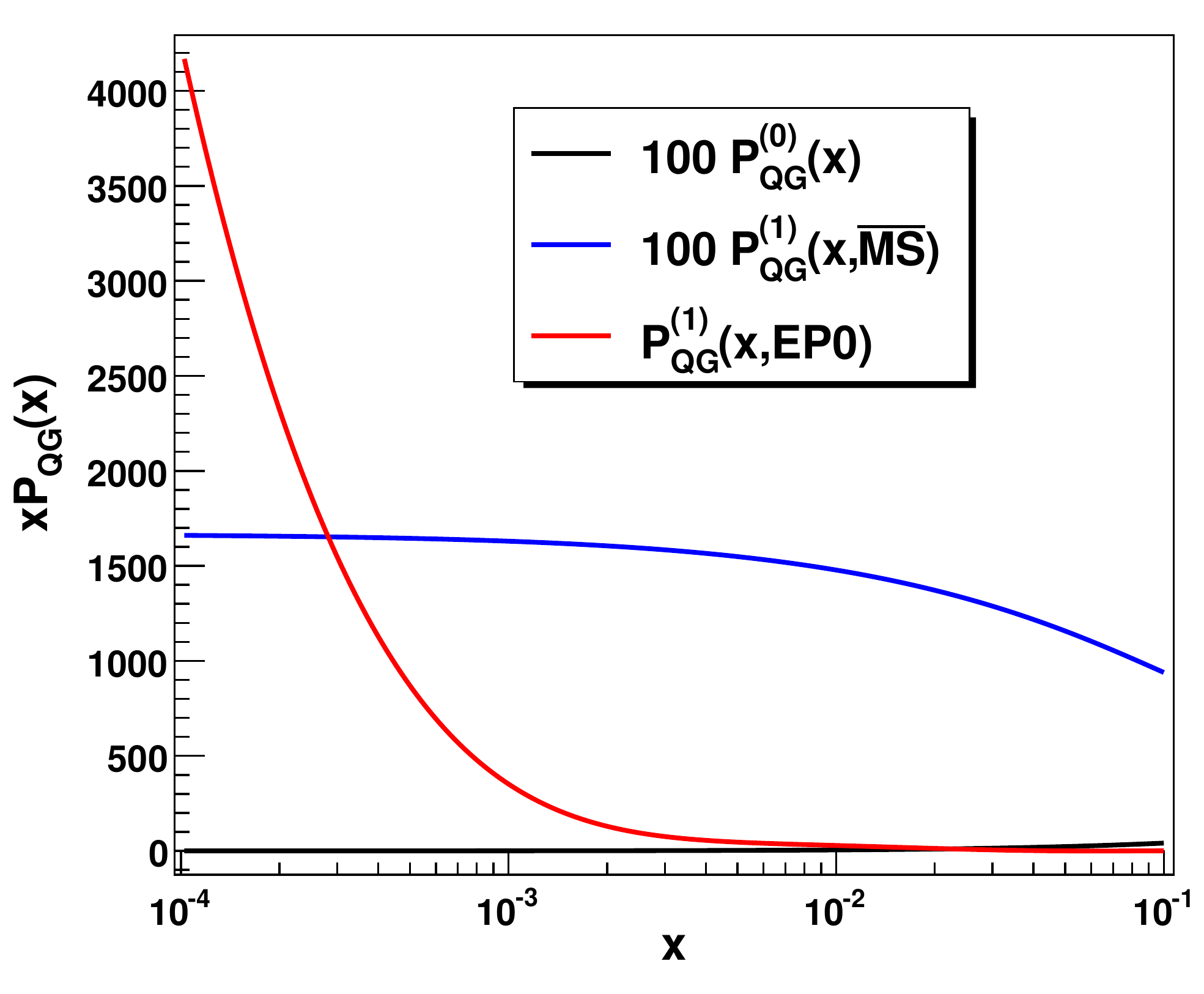}
  \includegraphics[width=0.4\textwidth,angle=90]{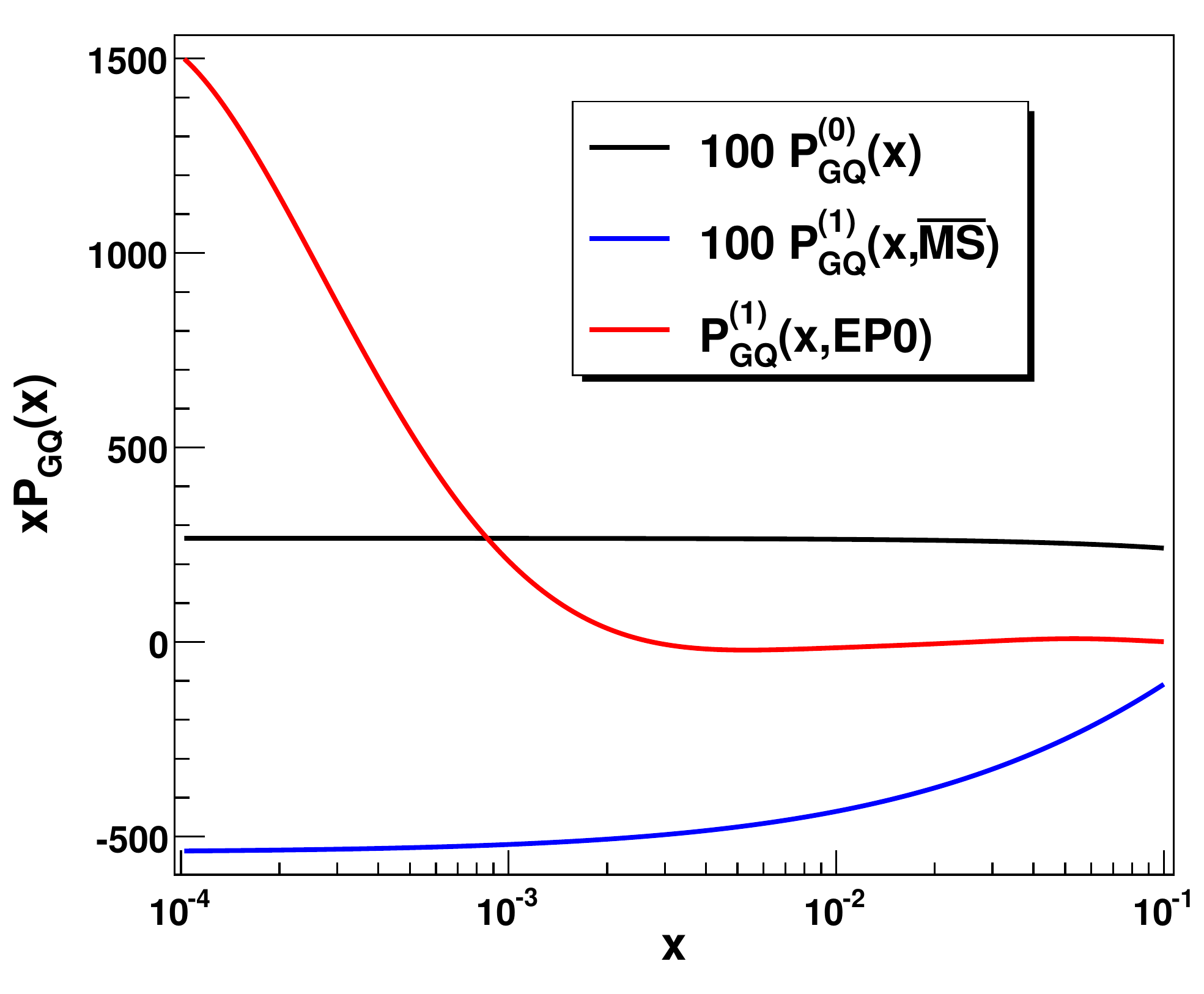}
  \includegraphics[width=0.4\textwidth,angle=90]{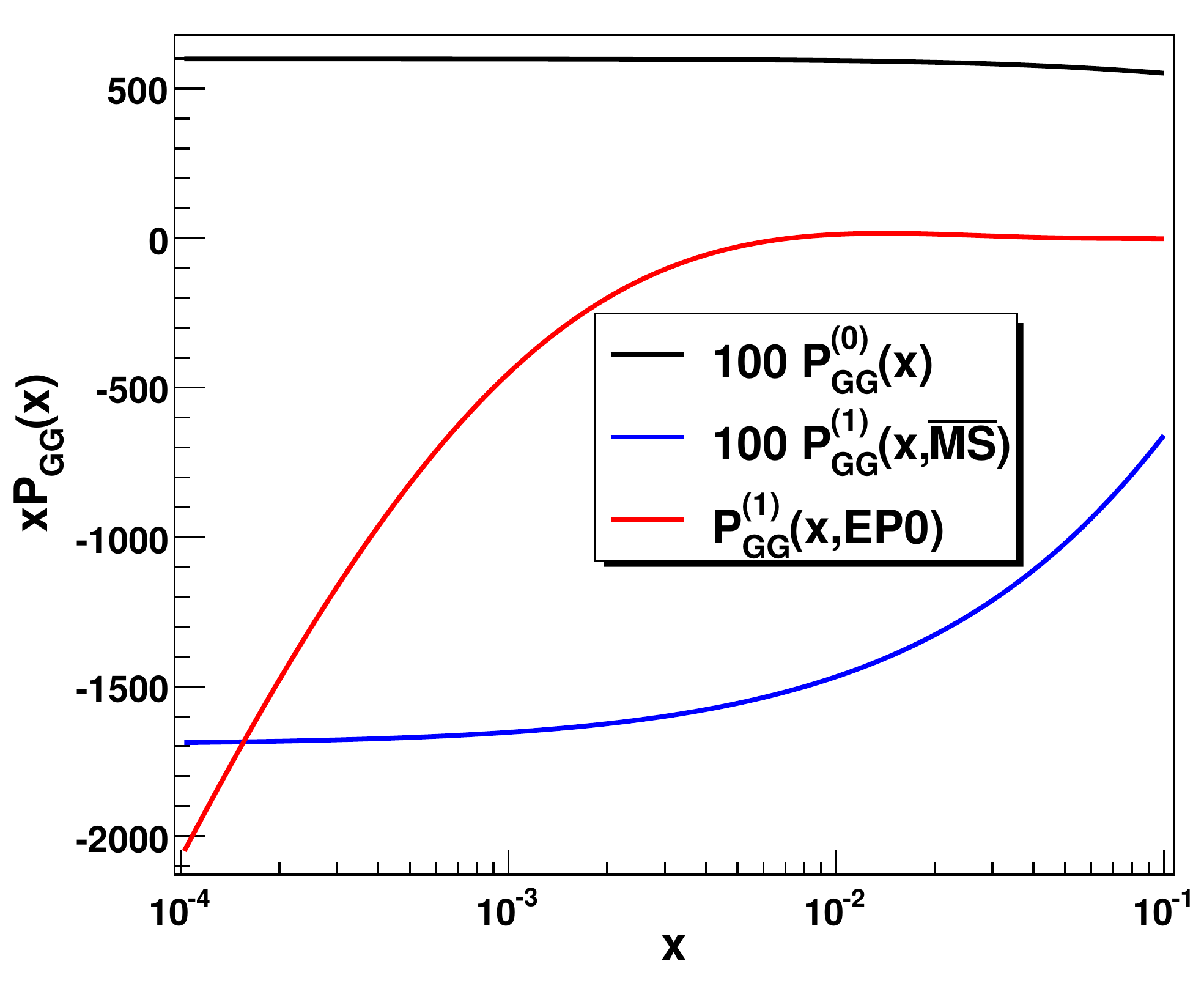}
  \caption{Comparison of singlet LO and NLO splitting functions at low $x$ for five massless
  quark flavours. The singlet NLO splitting functions are plotted for the $\overline{\rm MS}$
  and EP0 factorization scheme. Note that the singlet LO splitting functions and the singlet
  NLO splitting functions in the $\overline{\rm MS}$ factorization scheme are multiplied by
  100.}
  \label{figsplitfcelowx}
\end{figure}

The singlet NLO splitting functions in the EP0 factorization scheme are plotted in figures
\ref{figsplitfce} and \ref{figsplitfcelowx}. The figures show that the singlet NLO splitting
functions in the EP0 factorization scheme are not small in comparison with those in
the $\overline{\rm MS}$ factorization scheme \cite{nlosplitfuncsng}
and the singlet LO splitting functions. In the low $x$ region,
the singlet NLO splitting functions in the EP0 factorization scheme even strongly
dominate over the $\overline{\rm MS}$ and LO ones. Hence, the EP0 factorization scheme
cannot be regarded as a factorization scheme close to the ZERO factorization scheme. The
little influence of the singlet NLO splitting functions on the evolution of the parton
distribution functions of the proton in the EP0 factorization scheme is thus not a consequence
of the closeness of the singlet NLO splitting functions to zero, but it is a consequence of
their suitable shape.

Besides the EP0 factorization scheme, other factorization schemes in which the influence
of the NLO splitting functions on the evolution of the parton distribution functions of
the proton is almost negligible have been found. However, in all these factorization schemes
the behaviour of NLO splitting functions and parton distribution functions is similar to that
in the EP0 factorization scheme. If the functions $\mathbf{P}^{(1)}_k (x)$ in formula
(\ref{sfsnlofceform}) were chosen in such a way that they did not allow the singlet NLO
splitting functions $\mathbf{P}^{(1)} (x)$ to be large in absolute value in the low $x$ region,
then no NLO splitting functions with sufficiently small influence on the evolution of the parton
distribution functions of the proton in the corresponding factorization scheme were found.
It thus seems that in order that NLO splitting functions have little influence on the evolution
of the parton distribution functions of the proton, which is a necessary condition that
the factorization scheme specified by them be close to the ZERO factorization scheme, they
have to be large in the low $x$ region (provided that they satisfy the condition of practical
applicability (\ref{conpracapp})). But large values of NLO splitting functions are
in contradiction with the closeness to the ZERO factorization scheme. The results of searching
for an approximately ZERO factorization scheme based on the minimization of the difference
between $\mathbf{D}_0(x, M, {\rm FS}, M_{\rm S})$ and $\mathbf{D}(x, M, {\rm FS})$ thus
indicate that there is no factorization scheme which has no restrictions of its practical
applicability at the NLO and simultaneously is sufficiently close to the ZERO factorization
scheme. This conclusion is also confirmed by the results of a direct minimization of the singlet
NLO splitting functions $\mathbf{P}^{(1)} (x)$ given by formula (\ref{sfsnlofceform}) because
the NLO splitting functions that have been obtained by the direct minimization do not have a negligible
influcence on the evolution of the parton distribution functions of the proton, and therefore
the corresponding factorization schemes cannot be considered as close to the ZERO factorization
scheme.

%% file: approx_zero_fs3.tex
\section{Summary and conclusion}
\label{prtsumandconcl}

A factorization scheme suitable for current NLO Monte Carlo event generators
is the ZERO factorization scheme because its employment makes the combination
of formally LO initial state parton showers and NLO hard scattering
cross-sections consistent. But it has turned out \cite{kolar} that the ZERO
factorization scheme is not applicable in the full range of $x$ needed for
QCD phenomenology. The deficiency of current NLO Monte Carlo event generators
lying in combining formally LO initial state parton showers and NLO hard
scattering cross-sections can thus be removed by means of choosing a suitable
factorization scheme only in a certain kinematic region in which the ZERO
factorization scheme is applicable. Since this region covers, for instance,
the production of heavy objects, which is important in searches for new
physics, the use of the ZERO factorization scheme makes sense despite its
limited range of practical applicability. However, if we require applicability
in the full range relevant for QCD phenomenology, then the freedom in the
choice of the factorization scheme does not allow us to remove the mentioned
deficiency of current NLO Monte Carlo event generators because of the limited
practical applicability of the ZERO factorization scheme. Hence, we have
tried to find some factorization scheme which is applicable at the NLO in
the full range needed for QCD phenomenology and simultaneously sufficiently
close to the ZERO factorization scheme. Such a factorization scheme could
at least significantly reduce the mentioned deficiency of current NLO Monte
Carlo event generators.

We have found NLO splitting functions which satisfy the condition of practical
applicability (\ref{conpracapp}) and the influence of which on the evolution of
the parton distribution functions of the proton is almost negligible. However,
these NLO splitting functions are not small in comparison with those in the
$\overline{\rm MS}$ factorization scheme, and therefore the corresponding
factorization scheme cannot be regarded as a factorization scheme close to
the ZERO factorization scheme. Moreover, it has turned out that the condition
of practical applicability (\ref{conpracapp}) does not allow NLO splitting
functions to be significantly smaller (in the absolute value) than those in
the $\overline{\rm MS}$ factorization scheme. The deficiency of current NLO
Monte Carlo event generators lying in combining formally LO initial state
parton showers and NLO hard scattering cross-sections can thus not be significantly
reduced by means of choosing a suitable factorization scheme if we require
applicability in the full range relevant for QCD phenomenology.